# Spin-Orbit Logic with Magnetoelectric Switching: A Multi-Generation Scalable Charge Mediated Nonvolatile Spintronic Logic

Sasikanth Manipatruni[1], Dmitri E. Nikonov[1], Ramamoorthy Ramesh[2], Huichu Li[3] and Ian A. Young[1]

1. Components Research, Intel Corp., Hillsboro, OR 97124, USA
2. Department of Materials Science and Engineering, University of California, Berkeley, Berkeley, California 94720, USA
3. Intel Labs, Intel Corp., Santa Clara, CA 95054, USA

As nanoelectronics approaches the nanometer scale[1-4], a massive effort is underway to identify the next scalable logic technology beyond Complementary Metal Oxide Semiconductor (CMOS) transistor based computing[5-7]. Such computing technology needs to improve switching energy and delay at reduced dimensions[8], allow improved interconnects[9] and provide a complete logic/memory family. However, a viable beyond-CMOS logic technology has remained elusive. Here, we propose a scalable spintronic logic device which operates via spin-orbit transduction[10-16] combined with magneto-electric switching [17-20]. The Magneto-Electric Spin-orbit (MESO) logic enables a new paradigm to continue scaling of logic device performance to near thermodynamic limits for GHz logic[8,9,21] (100 kT switching energy at 100 ps delay). The MESO devices scale strongly and favorably with critical dimensions, showing a cubic dependence of switching energy on size $W$, $E_{meso} \propto W^3$, and a square dependence on voltage $V$, $E_{meso} \propto V^2$. This excellent scaling is obtained thanks to the properties of the spin-orbit effects (e.g. Inverse Spin Hall Effect (ISHE) and Inverse Rashba-Edelstein Effect (IREE)) and the dependence of capacitance on size. The operating voltages for these devices are predicted to be < 100 mV allowing a significant jump ahead of historic trends of scaling voltage with size and corresponding reduction of energy[22]. Interconnect resistance is a critical obstacle for scaling beyond 10nm dimensions. We show that MESO logic is amenable to operating with even highly resistive interconnects (100 µΩ.cm-1 mΩ.cm) which opens a possibility to use nanometallic (width << bulk electron mean free path) or doped semiconducting wires[23] for short range (< 1 µm) interconnects. A scalable, CMOS compatible, non-volatile logic family proposed in this paper may enable the next multi-generational scaling technology for computing devices.

In the past decade, the systematic pursuit of transistor scaling has been enabled by direct improvements to the carrier transport [e.g. 24-27] combined with superior electrostatic control [e.g. 1, 3, 4, 27-29]. In contrast to the pure dimensional scaling (Dennard scaling [8]), new transistor technology necessitated the use of strain [24], 3D electrostatic gate control (FinFET and nanowire transistors [27-29]), manipulation of the carrier effective mass, band structures and the gradual introduction of a variety of materials for interface and work function control, see e.g. [30]. Despite the successful scaling in size of transistors, voltage and frequency scaling have all but stagnated [31]. Further decrease of voltage has been hampered by the need for a high ratio (>$10^6$) of current between the 'on' and 'off' states, together with the Boltzmann limit of current control (60 mV/decade at room temperature). In the last ten years, a significant effort to invent, demonstrate, and benchmark possible beyond-CMOS devices got underway [6, 7, 32-44]. It includes alternative computing devices based on electron spin [32-39], electron tunneling [40-41], ferroelectrics [e.g 42], strain [e.g 43], and phase change [e.g 44].

Spin based logic is one of the leading options for the non-charge based computing [5,6] due to a) non-volatility when power to the circuit is turned off; b) higher logical efficiency (i.e., fewer devices required per combinatorial logic function) due to the use of majority gates [45]; c) memory-in-logic and logic-in-memory capability which enables a significant departure from von-Neumann computational architecture; d) possibility of integration with CMOS to create embedded caches [46]; e) amenability to neuromorphic [47] and stochastic computing [48]. In this paper, we propose a family of spin logic devices based on magneto-electric and spin-orbit effects that is energy efficient and addresses a prior shortcoming of all spintronic devices – the slow switching and interconnect speed – while it is also dimensionally scalable to meet the demands of next-generation computing.

Spin logic devices proposed so-far have been based on spin transfer torque [33-35], dipole coupled nanomagnets [36,37], spin wave [38,39], or magnetic domain walls [33] based devices. These traditional spin logic devices suffered from a) inefficiency of current-driven spin transfer torque switching; b) unfavorable scaling of magnetic dipole or field based switching; c) limited speed of propagation for magnetic domain walls; d) phase noise due to the low energy of spin waves. Intensive benchmarking efforts [6] have identified magneto-electric (ME) mediated spin devices to have switching energy comparable or smaller than that of charge based devices. However, an efficient way to cascade ME logic devices did not exist, because the prior art transduction from the spin to charge variable is limited due to the low values of tunneling magnetoresistance (TMR). Here we propose a spin logic device which uses magneto-electric switching in combination with an efficient spin to charge transduction mechanism.

### ***Spin to charge Conversion with spin-orbit effects:***

Firstly, we identify a scalable way to transduce the spin state of a nanomagnet to a charge state via spin-orbit effects [10-16] (such as the Interface Rashba-Edelstein Effect (IREE) and the bulk Spin Hall Effect (SHE)). In a recent discovery, it was shown that spin currents can be converted to charge currents preserving the information encoded in the spin polarization [14, 16, 49-54] (utilizing resonant spin pumping [14, 50-53] and in the quasi-static non-local spin valve configuration [54]). Fig. 1A shows how a current through a nanomagnet produces injection of spin-polarized electrons into a stack of materials with a high spin-orbit coupling coefficient (e.g. Bi/Ag [55], topological insulators [16, 56, 57], 2D materials [50], β-Ta [13], β-W [57], Pt [13], please see table 2). In Fig. 1A, when the nanomagnet's magnetization $\hat{m}$ is pointing in the $\hat{y}$ direction and the flow of injected spin current is $\vec{J_s} = J_s\hat{z}$, a charge current $\vec{I_c}$ is generated in the $\hat{x}$ direction. In Fig. 1B, when the nanomagnet is pointing in the $-\hat{y}$ direction and the flow of injected spin current is still $\vec{J_s} = J_s\hat{z}$, a charge current $\vec{I_c}$ is generated in the $-\hat{x}$ direction. Hence, the magnetization direction of the nanomagnet is transduced into the direction of electric current.

We derive the scaling law for spin to charge conversion using the properties of IREE and ISHE effects, in which the efficiency improves with reducing magnet width, a highly desirable scaling. The spin-orbit mechanism responsible for spin to charge conversion at the interface is described by the IREE. In the presence of Rashba spin-orbit interaction, the Hamiltonian of a 2D electron gas is:

$$H_R = \alpha_R (k \times \hat{z}) \bullet \vec{\sigma} \qquad (1)$$

where $\alpha_R = (k_{F+} - k_{F-})\hbar^2/2m$ is the Rashba coefficient, $k_{F+}, k_{F-}$ are Fermi vectors of the two spin-split bands, $\hat{z}$ is unit vector the normal to the interface, $\hat{\sigma}$ is the vector of the Pauli spin matrices and $k$ is the momentum of the electrons. In a simple model based on two Fermi contours in the Rashba electron gas (Fig 1C), the density of spin polarization along y-axis (Fig. 1) and the charge current density along x-axis (Fig. 1), can be related as [14, 15, 49]:

$$\delta s_{y\pm} = \pm \frac{m}{2e\hbar k_{F\pm}} j_{xc\pm}, \qquad (2)$$

which yields the relation between spin density and charge current in a 2D Rashba electron gas:

$$j_{Cx} = \frac{e\alpha_R}{\hbar} \langle \delta s \rangle_y \qquad (3)$$

This results in the generation of a charge current in the interconnect proportional to the spin current (Fig 1D, 1E). Relating linear charge current density j$_{Cx}$ (units of A/m) and areal spin current density j$_{sy}$ (spin current flowing along z-direction comprised of spins along y-direction, units of A/m$^2$) [15]

$$j_{Cx} = \frac{\alpha_R \tau_s}{\hbar} j_{sy} = \lambda_{IREE} j_{sy} \qquad (4)$$

where we assumed the relation between spin current and spin polarization is determined by the spin relaxation time as follows $j_s = e\langle \delta s \rangle/\tau_s$. For generality, we include the contribution from a bulk inverse spin Hall effect (with spin Hall angle θ$_{SHE}$ and bulk spin diffusion length λ$_{sf}$) which may co-exist with a IREE. We will refer to them together as inverse spin-orbit coupling (inverse SOC or ISOC) effects. In the presence of both effects the total charge current generated via spin-orbit effects is:

$$\vec{I}_c = \frac{1}{w}\left(\lambda_{IREE} + \Theta_{SHE}\lambda_{sf} \tanh\left(\frac{t}{2\lambda_{sf}}\right)\right)(\hat{\sigma} \times \vec{I}_s) = \frac{1}{w}\lambda'_{ISOC}(\hat{\sigma} \times \vec{I}_s) \qquad (5)$$

The spin-orbit interaction at the Ag/Bi interface (the inverse Rashba-Edelstein effect, IREE) and bulk ISHE produce a charge current in the horizontal direction [4]. We estimate that spin to charge conversion efficiency can be close to 100% for known material combinations. Both IREE and ISHE effects produce spin to charge current conversion with efficiency of ~0.1 with established SHE materials at 10 nm magnet width. For scaled nanomagnets (5 nm – 10 nm width) and exploratory ISOC/SHE materials such as $Bi_2Se_3$ and $MoS_2$, the spin to charge conversion efficiency can be close to 1.

### *Spin-Orbit Logic Device with Magneto-electric Input Signal Nodes:*

We detail a scheme of a spin-orbit logic device with magneto-electric nodes that utilize the strong scalability of spin to charge conversion combined with magneto-electric switching [17-20]. The building blocks of proposed MESO logic device comprise spin-orbit effects for spin to charge conversion at the output and magneto-electric switching for charge to spin conversion at the input and are shown in Fig. 2A and B. In Fig. 2A, when the input interconnect carries a positive current (positive carriers flowing in **+x** direction), an electric field is set up in the magneto-electric capacitor in the **-z** direction. The resulting magneto-electric field ($H_{ME}$) switches the nanomagnet to the **-y** direction [57, 17-20]. Due to the SOC spin to charge transduction (Fig. 1B), a charge current is generated at the output in the **-x** direction. The macro-magnetic simulation of the MESO inverter transfer function with electric and magnetic hysteresis, that is advantageous for noise rejection, is shown in Fig. 2C and D. Large-signal gain of the output current (the ratio of $I_{out}/I_{in}$) is generated and controlled by the supply current ($I_{supply}$). Small-signal gain (the derivative of $I_{out}$ relative to $I_{in}$) of the device during switching can be seen in Fig. 2D. The charge current carried by the interconnect, creates a voltage on the capacitor comprised of magnetoelectric material (such as $BiFeO_3$ or $Cr_2O_3$, please see table 2) in contact with an output nanomagnet which serves as one of the plates of the capacitor. Typical magnetoelectric structures include either intrinsic multiferroics or composite multiferroic structures. As the charge accumulates on the magnetoelectric capacitor, a strong magnetoelectric interaction causes the switching of magnetization in the output nanomagnet.

The innovative impact of MESO logic is realized with the output charge mediated interconnect at its output, shown in Fig. 2E. The previous logic stage injects charge current ($I_C$) into the interconnect which charges the magneto-electric capacitor at the input of the next MESO logic stage. Charge interconnect enables long distance signal transmission that does not suffer from the limitations such as spin diffusion length [6] or domain wall propagation lengths [6]. The operation of the charge interconnect does not rely of the spin coherence of the carriers at the magnetoelectric detection node. Charge interconnects also significantly enhance the suitability of MESO logic for 3D integration. Spin based interconnects would also be incompatible with metallic layers used in forming the 3D vias in CMOS electronics (such as W, Ta). We now show the transient operation of the device using vector spin circuit theory [35, 58]. A spin equivalent circuit which models the scalar charge current, and the spin current with three component Cartesian vectors of spin currents ($I_{sx}$, $I_{sy}$, $I_{sz}$) is shown in Fig. 3A. Please see supplementary and method section.

***Dimensional Scaling and voltage scaling laws for MESO logic:***

The proposed spin-orbit logic with magneto-electric nodes exhibits an excellent scaling law where the energy to switch the device improves with dimensional scaling of the device. The scalability of the device can be attributed to: a) improvement in the spin to charge conversion efficiency ($\eta_{SOC}$) with reduction in the nanomagnet width; b) reduction in capacitance of the magneto-electric node with areal scaling. The energy to switch a single MESO logic unit is given by (Please see supplementary section F and G):

$$E_{MESO} = E_{CME} + E_{IC} + E_{ISOC} + E_{RT} + E_{SG} = C_{me} V_{me}^2 \left(1 + \alpha \frac{W}{\lambda'_{SOC}}\right)$$

Where $\lambda'_{ISOC}$ is the effective SOC conversion length (comprehending the bulk spin-orbit effect contribution):

$$\lambda'_{ISOC} = \left(\lambda_{IREE} + \Theta_{SHE} \lambda_{sf} \tanh\left(\frac{t}{2\lambda_{sf}}\right)\right)$$

Figure 4A, shows the strong cubic scaling of MESO energy where the energy reduces by 8X for every 2X reduction in feature size. The excellent scalability of MESO logic allows for switching energy of the MESO logic to approach $10^{-18}$ J per bit (Fig. 4A). Magneto-electric switching also allows for a strong square law voltage scaling in energy per bit. Figure 4B, shows a combination of scaling in energy/switching via voltage and effective IREE length, for e.g effective $\lambda_{IREE}$ of 0.6 nm [e.g 50] and switching voltage of 200 mV [e.g 65] allow < 10 aJ function. Further lowering of the switching voltages can enable quadratic energy scaling. Please see section F of the supplementary for detailed energy calculations.

### *Dynamic variations of Magneto-electric spin-orbit logic:*

We show that MESO logic is highly tolerant to stochastic dynamic variations in switching inherent to operating at attojoule (aJ) switching levels. This is also in sharp contrast to traditional spin logic devices. The Langevin thermal noise in magnetic devices produces switching speed (delay) variations that can be a limiting factor for logic applications. In particular, spin torque switching has been shown to have a long tail in its switching probability vs. switching time. This forces using longer switching current pulses and results in excessive switching energy. In Fig. 4D, we compare the stochastic switching nature of the MESO logic with highly scaled spin torque logic devices (called *all-spin-logic, ASL*) employing Heusler alloys [59, 35] and perpendicular magnetic anisotropy [35]. The ASL and MESO devices are modeled using vector spin circuit theory (please see supplementary). The nanomagnet's Langevin noise is treated using Monte Carlo simulation of stochastic terms in the micromagnetic Landau-Lifshitz-Gilbert (LLG) equations. We calibrated the stochastic LLG vs. Fokker-Planck equation for the probability distribution of the angle of magnetization. Magneto-electric switching modifies the magnetic anisotropy energy landscape allowing for a fundamentally different characteristic of switching error distributions. Fig. 4D compares the switching error function for MESO, STT-MRAM, ASL-Heusler [60] and ASL with perpendicular magnetic anisotropy (PMA) to show that MESO logic can enable a switching time in the 100 ps range with error rates of less than $10^{-14}$.

The proposed MESO logic device enables competitive energy-delay performance compared to leading beyond-CMOS device options while also allowing non-volatility. We compare the MESO logic device with other leading beyond-CMOS options in a higher level benchmark circuit – a 32-bit adder (Fig. 4F, please see supplementary). The MESO logic device enables significant speed improvement compared to other spintronic devices due to the inherent speed of magneto-electric switching [17-20] along with shorter delay in interconnects. MESO also enables an energy reduction compared to CMOS logic operating at very low power (0.3V supply voltage) due to the ability to switch at even lower bias supply voltages (0.1V). The speed of MESO logic units is comparable to the low power, low leakage CMOS devices (0.3V supply voltage). We note that at low logic activity factor and intermittent usage, the non-volatility of the MESO device can enable further advantages compared to CMOS by eliminating the standby power dissipation and enabling instantly on operation from standby.

### *Interconnect Scaling beyond 10 nm: electrically mediated charge interconnects for MESO:*

Scalability of interconnects for CMOS has emerged as a major limitation when the width of the electrical wires have reached < 20nm [9]. While patterning techniques (computational lithography, inverse lithography and extreme ultra violet lithography) are available to reduce the dimensions further, the electrical resistivity (and spin resistivity) will be increased significantly. Experimental data for highly scaled interconnects show that the resistivity of electrical wires increases following the Mayadas-Shatzkes (MS) scaling law as [61] $\rho = \rho_0 \left(1 + \frac{3\lambda_{ebulk}}{8t}\left(1 + \frac{p}{2}\right) + \frac{3\lambda_{ebulk}}{2D}\left(\frac{R}{1-R}\right)\right)$ where $\rho_0$ is bulk resistivity, $\lambda_{ebulk}$ is the electron mean free path, *p, and R* is the specularity, reflection parameter from grain boundaries) as critical interconnect dimensions approach the electron mean free path. A second scaling issue with electrical interconnects is the capacitance (*C*) per unit length. The intrinsic speed of the long interconnects is limited by capacitive *RC* charging time parameter. The capacitance of the wires increases for tightly spaced interconnects requiring the adoption of high porosity low dielectric constant materials as interconnect dielectrics, which are inherently limited to a narrow range of dielectric constants $1.3 < \epsilon_{dielectic} < 2.5$. Spin interconnects (and magnetic domain wall) interconnects also suffer from limited spin coherence length (and domain wall propagation length) as nanowire dimensions approach 20 nm. Hence, it is of great interest to demonstrate a logic technology compatible with high resistivity and high capacitance interconnects.

We show that MESO logic can tolerate the use of high resistivity nanometallic interconnects with resistivity > 100 μm.cm and capacitance > 100 aF/μm. We show the interconnect scalability of electrically mediated charge interconnect used in MESO logic in Fig. 5. Fig. 5A shows the impact on switching time as a function of interconnect resistance for a 45nm interconnect length. The MESO interconnect tolerates increases in line resistance to 1 kΩ/micron (Fig. 5A) and increases in interconnect capacitance to 200 aF/micron. This would represent a significant relaxation in the demands placed on nanometallic interconnects compared to CMOS. The switching speed of the MESO interconnect scales linearly with interconnect length up to 1000 nm using an interconnect with the line resistance of 100 Ω/μm and the capacitance of 100 aF/μm. This is in contrast to spin interconnects where the switching speed degraded as $e^{-x}/L_{sf}$. Hence, MESO logic alleviates the traditional problem of interconnects for spin logic and as well as allows for continued scaling of metallic and semiconducting wires potentially to nanometer widths [23].

### *Material requirements for 10 aJ class MESO logic:*

We describe the material scaling path for 10 aJ class MESO logic scalable to 10 nm critical dimensions. Nominal material targets for 10 nm magnetic dimensions with practical material parameters are listed in Table 1. We considered experimentally shown material properties for inverse Rashba Edelstein parameter ($\lambda_{IREE}$ ~ 0.6 nm shown for $MoS_2$ [50], α-Sn [63] and Bi/Ag surface alloy [14]). Large signal magneto-electric (ME) coefficient of 10/c from ME switching [18] and low coercive voltages were enabled via tetragonality tuning and chemical substitution of multiferroics [64] with thickness scalability to 20 nm [65]. The output resistance of the current source of the spin to charge conversion is a critical parameter impacting the driving ability of the MESO logic device (high source resistance is preferred for a current source). Please see section H of the supplementary and Figure S9. Low interconnect resistivity requirement is significantly relaxed due to the low voltage, low current operation of charge mediated magneto-electric logic. This is reflected in the resistivity targets of 4-200 µΩ.cm which are comparable to resistivity in scaled metal wires. Electromigration of the metal interconnect imposes a challenging limit to the scalability of switching speed by limiting the peak performance in computing circuits. MESO logic relaxes the electron-migration requirements to 25 MA/cm$^2$, significantly below the Belch limit for electron-migration of interconnect metal candidates [66].

## *Material scaling path for MESO logic*

We describe the material scaling path for 10 aJ class MESO logic scalable to < 10 nm critical dimensions or device density beyond 10$^{10}$/cm$^2$. We describe an array of possible material choices for MESO logic potentially suitable for its large scale integration. The four classes of materials for scalable MESO logic are a) Spin-orbit Coupling (SOC) materials for spin to charge conversion (ISOC) b) Magneto-electrics for charge to magnetic conversion c) Interconnects scalable to nanoscale widths d) Nanomagnet materials. SOC materials can be comprised of a) high spin-orbit coupling metals and their super lattices (Bi/Ag [14], Pt, β-Ta [13], β-W[57], CuBi, CuIr, Bi$_2$O$_3$ [67], W(O) [68], Ag/Sb [53]) b) topological materials] (Bi$_{1.5}$Sb$_{0.5}$Te$_{1.7}$Se$_{1.3}$ [16], Bi$_2$Se$_3$ [56], Sn-Bi$_2$Te$_2$Se, $\alpha$-Sn [63] ) and their super lattices [69] c) Transition Metal di-chalcogenides with large spin-orbit effects (MoS$_2$[50], WSe$_2$). The magneto-electric materials can be comprised of a) Multiferroics with coupling of anti-ferromagnetic order and ferroelectric order (BiFeO$_3$, LaBiFeO$_3$, LuFeO$_3$ [70] and super-lattices) b) Magnetostrictive materials (Fe$_3$Ga [75], Tb$_x$Dy$_{1-x}$Fe$_2$ [76], FeRh [77]) c) Electrically tuned exchange mediated magneto-electrics (Cr$_2$O$_3$ [20]). The interconnect options scalable to < 10 nm critical width dimensions can be transition metal based (Cu, Ag, Co, Al, Ru) or their semiconductor alloys (poly-Si, NiSi, CoSi, NiGe, TiSi) combined with low interconnect capacitance materials (SiO$_2$, SiN, SiCOH, Polymers). The nanomagnetic materials can be ferromagnets/ferrimagnets (Co, Fe, Ni, CoFe, NiFe, $X_2YZ$, XYZ alloys e.g. $Co_2FeAl$, $Mn_3Ga$), in which a wide range of saturation magnetization and magnetic anisotropy are feasible to meet the dimensionality and retention requirements. In spite of the wide range of available materials, significant material development is required to a) improve the material interfaces for integrated devices b) improve the range of operating temperature c) improve the processing temperature compatibility.

### *CMOS compatibility, memory and control logic:*

The proposed magneto-electric logic has several desirable features that are compatible with CMOS nanoelectronics. First, MESO can be integrated in the backend of the CMOS process (i.e., between the interconnect layers) and allowing for CMOS devices to be used for clocking control and power features (Fig. 6). Second, MESO contains a feasible "logic-compatible speed" embedded memory (known as large signal memory, commonly implemented with SRAM) making it usable as an on-chip non-volatile memory. Third, the MESO device can allow stacking of several layers of magnetic logic in a 3D architecture. Fourth, since the state variable of the interconnect between MESO gates is charge, MESO logic can readily interface with CMOS circuitry to implement the clocking control, and power delivery.

### *Conclusion:*

In conclusion, we propose a scalable beyond-CMOS spintronic logic device with non-volatility and with high speed energy-efficient charge based interconnect. The proposed device allows for a) continued scaling in energy per operation aJ/switching at 100 ps switching speed; b) improved scalability for interconnects due to its insensitivity to interconnect resistivity up to 1 mΩ.cm; c) reduced operating voltage down to 100 mV and even potentially lower; d) improved stochastic performance compared to spin torque logic devices with "logic class" error rates ($<10^{-14}$); e) a path to seamless integration with CMOS structurally as well as for processing charge based information. The ability to transition to a beyond CMOS device with an advantageous method of scaling utilizing novel magnetic materials, high resistivity but still highly reliability interconnects, employing majority logic, and utilizing non-volatility can open up a potentially new technology paradigm for improving energy efficiency in beyond CMOS computing devices.

# METHODS

## A. Vector Spin Circuit Modeling of MESO Logic:

We verify the functionality of MESO spin logic using an equivalent spin circuit model, which comprehends magnetization dynamics of the nanomagnet, vector spin injection, spin to charge transduction and the magneto-electric switching. The equivalent circuit model is based on vector spin circuit theory [S1-4] (magneto-electric circuit analysis). The spin to charge conversion is modeled as a Spin Current Dependent Charge Current Source (SCDCCS) governed by the ISOC transduction:

$$I_{c\,\text{int}} = \frac{1}{w}\left(\lambda_{IREE} + \Theta_{SHE}\lambda_{sf}\tanh\left(\frac{t}{2\lambda_{sf}}\right)\right)(\hat{\sigma}\times\vec{I}_s)\bullet\hat{y}$$

An intrinsic resistance of the ISOC current source is assumed based on the conductivity of the interconnect and the ISOC conversion layers. The nanomagnet is connected to a control transistor operating as a switched source of current shared among several MESO devices. We have also included the resistance and capacitance parasitics of the ground contact. Conductance across the magnet to the spin injection layer (B) is modeled as a 4 component spin conductance which relates the applied charge voltage and spin voltages to the injected charge and spin currents [35]. The 4-component current (comprising the charge current and the 3 Cartesian vector components of the spin current polarization) injected at the NM-FM interface are given by,

$$\begin{bmatrix}I_c\\I_{sx}\\I_{sy}\\I_{sz}\end{bmatrix} = R^{-1}(\hat{m})\begin{bmatrix}G_{11} & \alpha G_{11} & 0 & 0\\ \alpha G_{11} & G_{11} & 0 & 0\\ 0 & 0 & G_{SL} & G_{FL}\\ 0 & 0 & -G_{FL} & G_{SL}\end{bmatrix}R(\hat{m})\begin{bmatrix}V_N - V_F\\V_{sx}\\V_{sy}\\V_{sz}\end{bmatrix} \quad (3)$$

Where **R** is the rotation matrix to account for the magnetization direction of the nanomagnet, $I_{si}$ and $V_{si}$ are the vector spin currents and voltages). Please see supplementary for a detailed explanation of the modeling.

## B. Energy of ME switching

If we choose the following parameters of the magnetoelectric capacitor: thickness $t_{ME} = 5nm$, dielectric constant $\varepsilon = 500$, area $A = 60nm \times 20nm$, then the capacitance is

$$C = \frac{\varepsilon \varepsilon_0 A}{t_{ME}} \approx 1 fF$$

Demonstrated values of the magnetoelectric coefficient are up to $\alpha_{ME} \sim 10/c$, where the speed of light is $c$. This translates to the effective magnetic field exerted on the nanomagnet

$$B_{ME} = \alpha_{ME} E = \frac{\alpha_{ME} V_{ISHE}}{t_{ME}} \sim 0.06T$$

The charge on the capacitor $Q = 1fF * 10mV = 10aC$, and the time to fully charge it to the induced voltage is $t_d = 10Q/I_d \sim 1ps$. If the driving signal voltage is $V_d = 100mV$, then the switching energy is

$$E_{sw} \sim 100mV * 10\mu A * 10ps \sim 10aJ, \tag{6}$$

which is comparable to the switching energy of CMOS transistors [2]. Note that the time to switch magnetization remains much longer than the charging time and is determined by the magnetization precession rate. The micro-magnetic simulations of magneto-electric switching predict this time to be $t_{sw} \sim 100ps$.

**C. Stochastic behavior of Magneto-electric switching Vs Spin Torque Switching**

We modeled the magnetization dynamics of the nanomagnet using a) Landau-Lifshitz-Gilbert (LLG) equation [11] b) Fokker-Planck equations [12, 13]. The phenomenological equation describing the dynamics of nanomagnet with a magnetic moment unit vector ($\hat{m}$), the modified Landau-Lifshitz-Gilbert (LLG) equation was used for Monte Carlo simulations (see Table S1 for parameters). We use Fokker-Planck equation for uniaxial anisotropy parameterized with angle of the magnetization, validated with Monte Carlo simulations of the nanomagnets. Please see supplementary for a detailed explanation of the stochastic modeling.

**D. Uniform benchmarking to beyond CMOS logic options**

We adopted the uniform benchmarking method developed to compare the beyond CMOS options. This method comprehends the impact of material improvements, device topology and impact of interconnects. The model is adopted by the beyond CMOS researchers for comparing  a) Spin torque (STT-DW: Spin Transfer Torque  Domain Wall, ASLD: All Spin Logic Device, CSL (Charge Spin Logic), STO logic (Spin Torque Oscillator logic)) b) Dipole field (NML, Nano-Magnetic Logic) c) Magneto-electric (MESO, SMG, Spin Majority Gate, SWD, Spin Wave Device). We evaluated digital logic benchmarking circuits [20, 21], a fanout-4 inverter, a 2-input NAND, and a 32-bit ripple-carry adders to compare the MESO logic with leading beyond CMOS logic options. Please see supplementary for a detailed explanation of the benchmarking.

**E. Complete logic family and State elements**

The proposed device family readily extends to a general purpose computing state machine. A state machine and complete Boolean logic family are the pre-requisites for a Turing Machine [81]. Majority logic operation can be readily shown since the input to a capacitive node is additive to the charge currents converging at the node via Kirchhoff law. Spin logic devices with multiple switching inputs (domain wall/spin wave/spin current) have been shown to allow majority logic [5, 6] and spin state machine [82]. Combined with a Randomly Accessible Memory (RAM) (Fig. 2C, D, Fig. 6) a state machine enables general purpose computing.

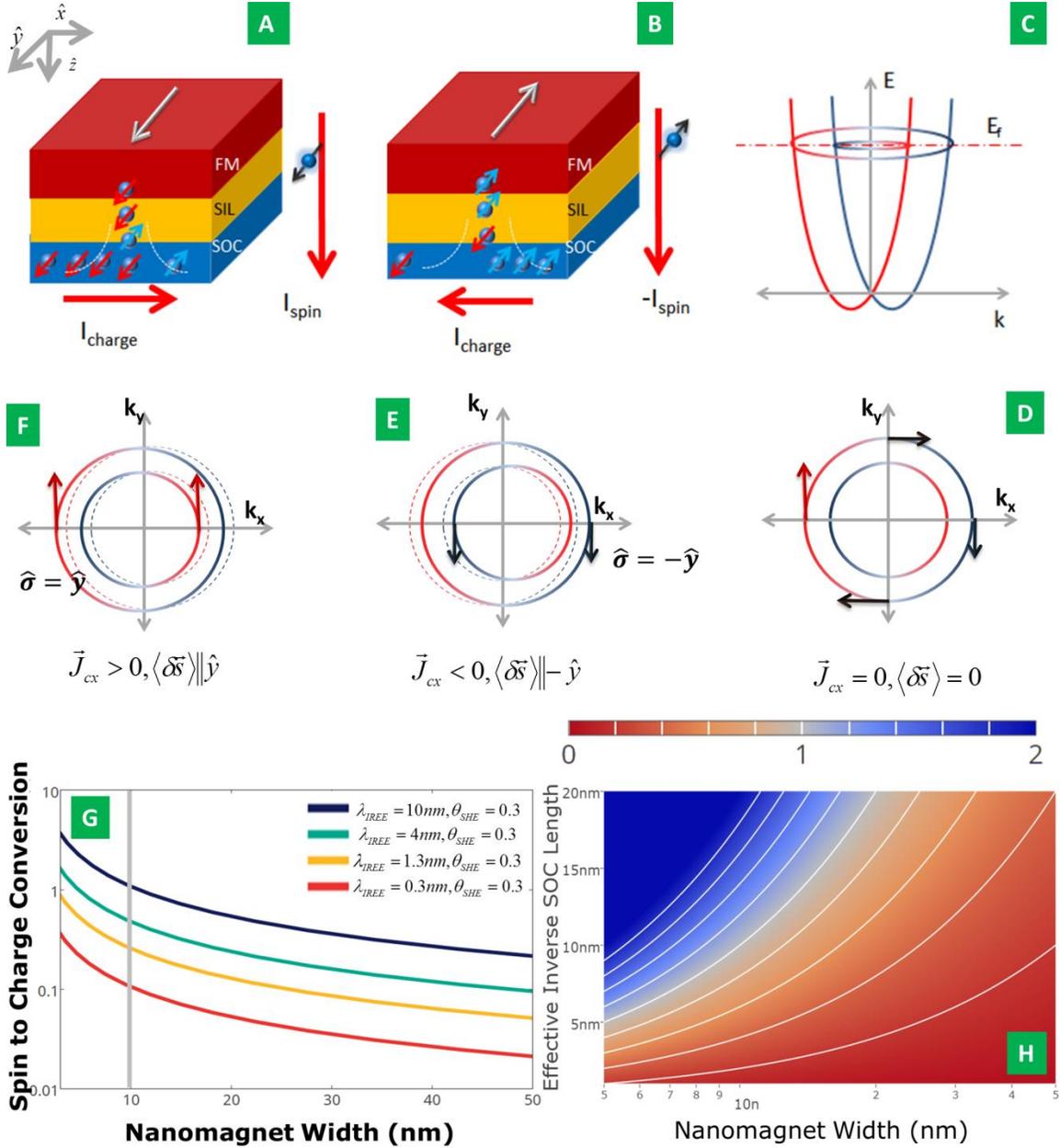

*Figure. 1 Scaling of spin to charge conversion via spin-orbit effects (Inverse Rashba Edelstein Effect and Inverse Spin Hall Effect). A) Injection of electron current with spin polarization along (+y) produces a charge current response in the lateral direction (+x). B) Reversing the direction of the magnet reverses the charge current response. C) Electronic structure of interface Rashba states with spin-split dispersion of 2DEG. D) Typical Fermi contours at the Fermi-level, electron momentum and spin state are related for a given band. Electrical current produces a spin flow and vice versa. E) Fermi-contours during spin injection along -y direction, producing a net charge flow in -x direction. F) Fermi-contours during spin injection along +y direction, producing a net charge flow in x direction. G) Spin to charge conversion efficiency as a function of magnet width. H) Spin to charge conversion efficiency as a function of Effective $\lambda_{ISOC}$ and nanomagnet width showing the feasibility of reaching high conversion efficiency for various combinations*

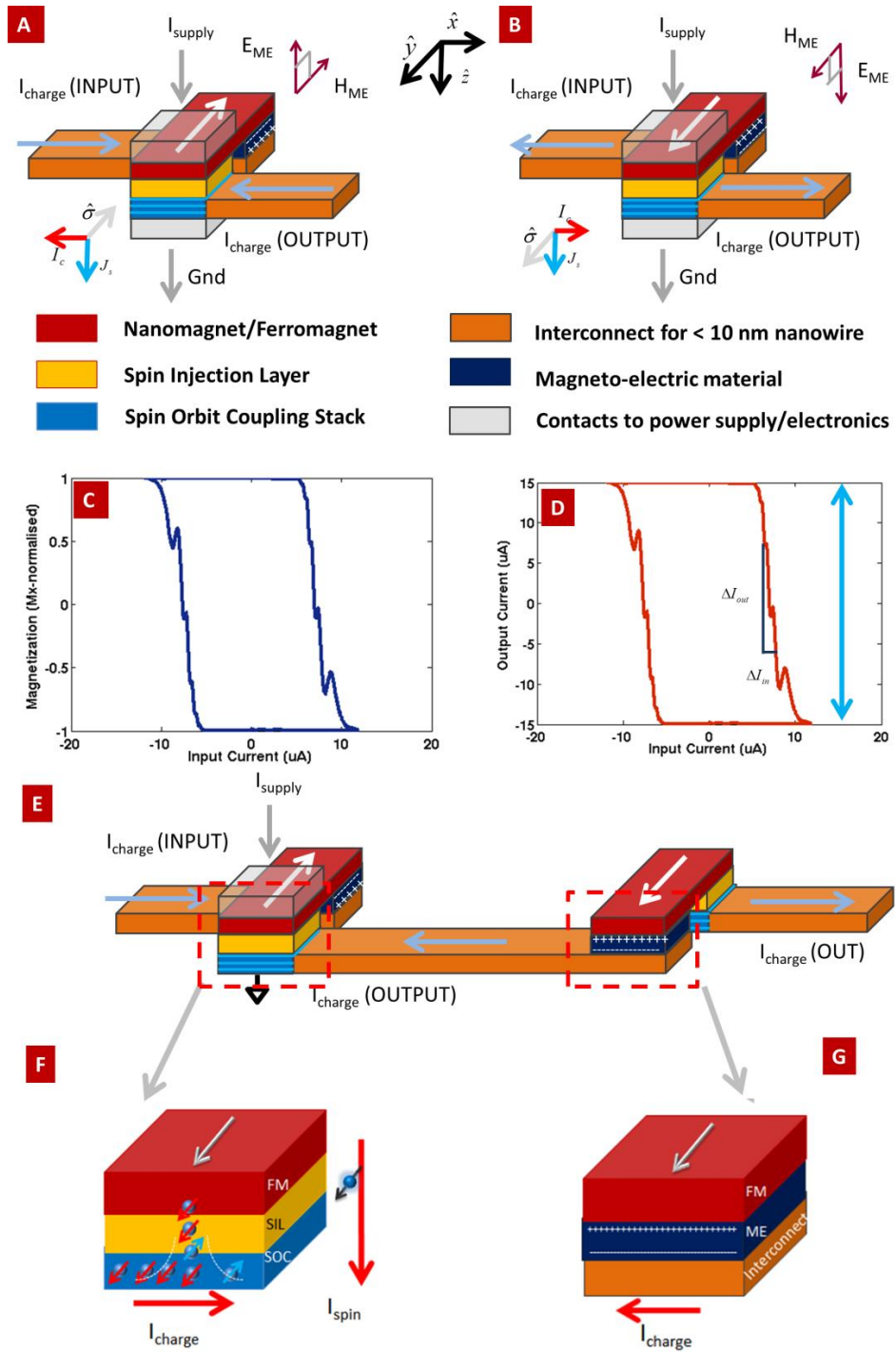

*Figure 2. Charge in - charge out inverting logic unit for MESO. A) Positive charge current (1) to negative charge current (0) conversion. B) Negative charge current (0) to positive charge current (1) conversion. C) DC transfer function for magnetic state with input current. D) DC transfer function for output current vs input current. E) The interconnect between two cascaded MESO inverters. F) Transduction from magnetic state variable to interconnect charge current with IREE and ISHE effects. G) Transduction of electrical signal $I_c$ on the interconnect to magnetic state.*

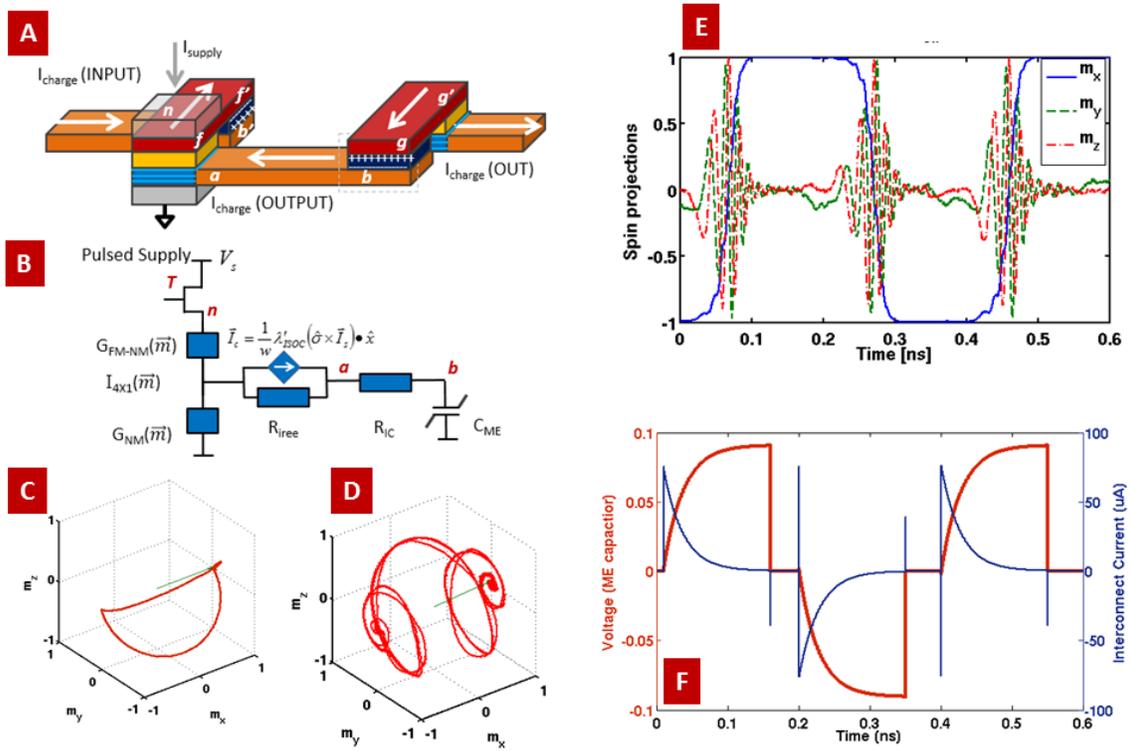

*Figure 3 Transient vector spin modelling of the MESO device. A), B) Spin circuit model using vector spin circuit models. C), D) Input and output magnetic response. E) Response of the output magnet for input magnetization. F) Voltage and current applied on the ME capacitor and interconnect. (Please see supplementary for detailed circuit simulation)*

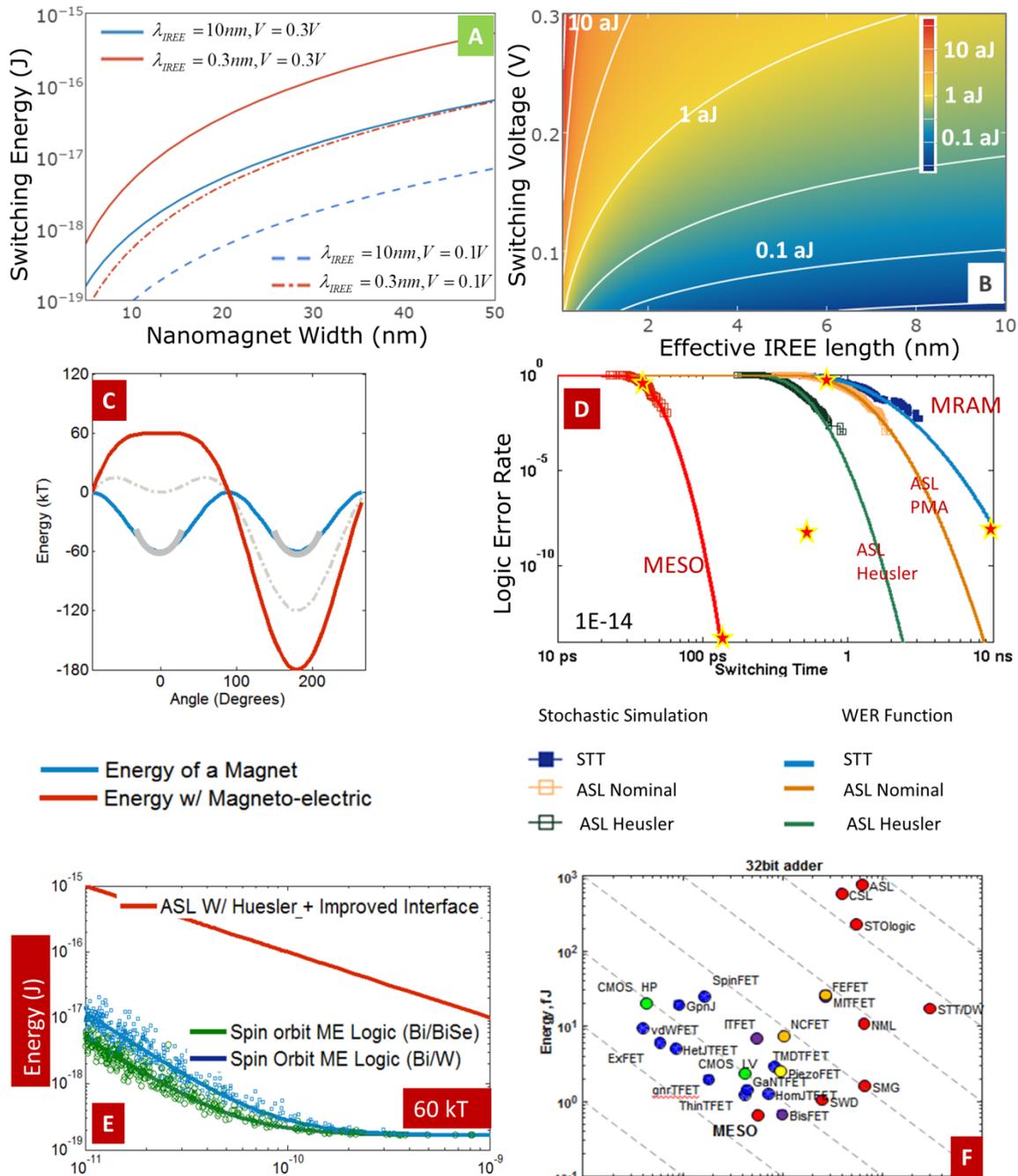

*Fig. 4. Dimensional Scaling of MESO logic showing accelerating gains in switching energy as the devices get smaller. A) Cubic scaling with magnet width. B) Scaling of energy with voltage and effective IREE length. C) Shaping of the energy barrier in magneto-electric vs spin torque switching. D) Write error rate statistics of MESO logic vs ASL and STT_MRAM. E) Energy vs. delay of a single MESO gate. F) Comparison of the MESO logic with leading Beyond-CMOS logic technologies Energy vs delay for a 32bit adder.*

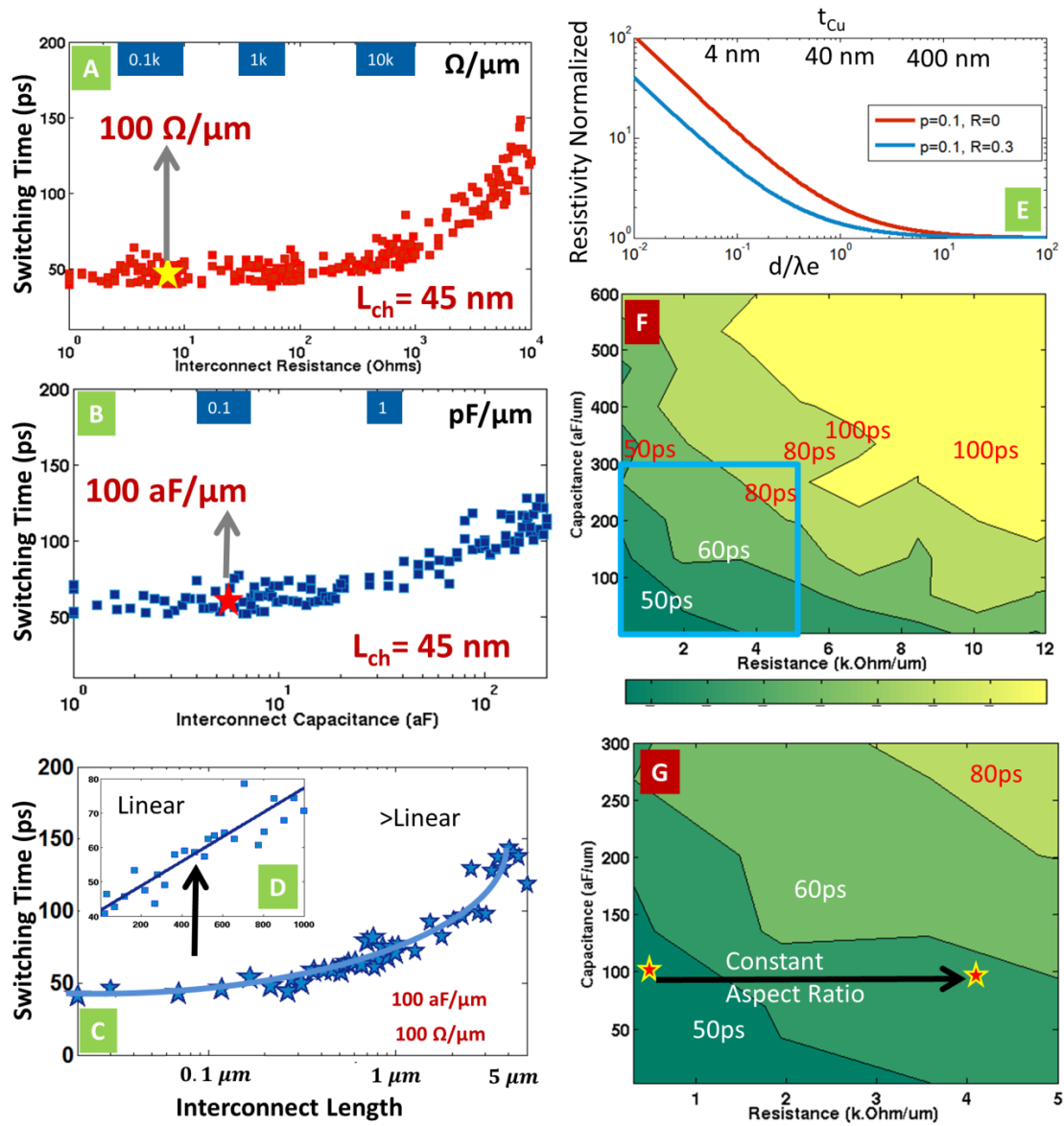

*Figure 5. Interconnect scalability for MESO Logic. A) Switching delay scaling with interconnect resistance for a 45 nm interconnect (also expressed in resistance per unit length). B) Delay scaling with interconnect capacitance per unit length. C) Switching delay vs interconnect length showing linear delay increase with length. D) Linear delay scaling with length of interconnect (nm). E) Scaling of resistivity of electrical interconnects showing the impending problem with nanometallic wires scaled to < 10 nm width. F) Combined sensitivity of MESO performance to interconnect capacitance and delay showing minimal performance impact G) Impact on delay with constant aspect ratio scaling from 100 Ω/µm to 4 kΩ/µm interconnects. .*

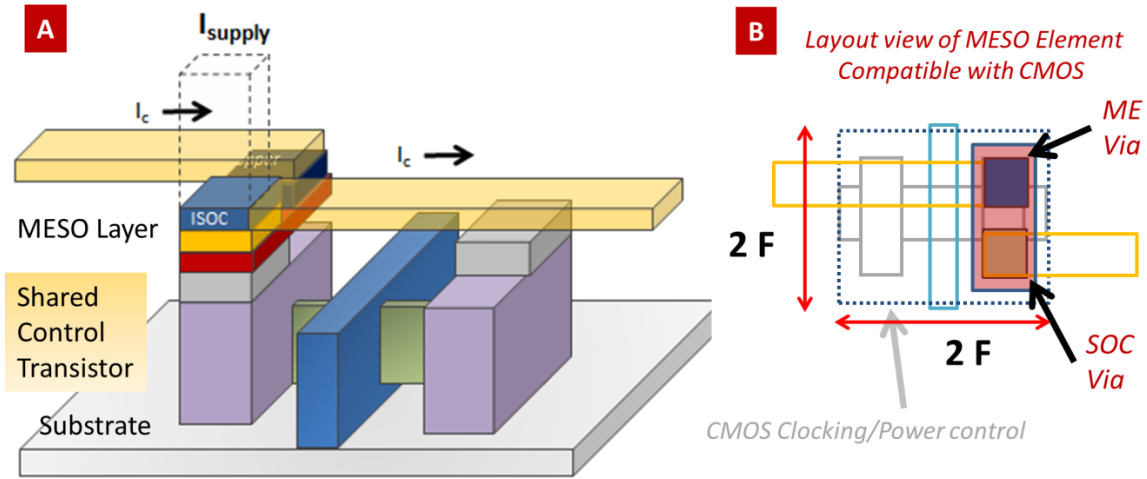

*Figure 6. Structural compatibility of the MESO logic with nanoelectronic CMOS. One possible integration scheme of CMOS with MESO logic is shown where scaled 3D FinFETs are monolithically integrated with MESO. MESO materials are formed near the 'Front End' transistors where higher temperature processing is permitted. Each transistor acts as a clock/power source for a number of MESO devices due to the low drive current/voltage needs of MESO. A) CMOS + MESO logic integration with MESO devices integrated into the interconnect layer of scaled CMOS. B) A top view of a compact cell comprising of a control/power transistor and MESO logic. Control transistor is shared among several MESO devices.*

| | Device Figure of Merit | Nominal Target | Material Figures of Merit | Nominal Target (for 1-10 aJ/switch) |
|---|---|---|---|---|
| **SOC Materials** | Spin to Charge conversion ($I_c/I_s$) | > 50 % | $\lambda_{IREE}$ | > 5 nm [50] |
| | Source Resistance | > 10 k Ohms | Resistivity | > 10 mΩ.cm [50, 56, 57] |
| **Magneto-electrics** | Dimensions | < 10 X 10 X 10 $nm^3$ | | |
| | Equivalent Capacitance/Area | < 100 fF/$\mu m^2$ | Charge/Area (for 1-10 aJ target) | 0.5-5 $\mu C/Cm^2$ [64, 70, 74] |
| | Switching Voltage | 0.1 – 0.3 V | Coercive Field | 100- 500 kV/cm [64] |
| | | | Magneto-electric Coeff. ($\alpha_{ME}$) | 10/C [18, 20, 64] |
| **Interconnect** | Resistance/ Length | 0.1-5 kΩ/μm | Resistivity | 4 -200 $\mu\Omega$.cm [61] @ 10 nm width |
| | Capacitance/ Length | 10-100 aF/μm | ILD Dielectric constant | 1-10 [61] |
| | Peak Currents | 10-100 μA/Magnet | Electromigration Limit | > 25 $MA/cm^2$ [61] @ 10 nm width |
| **Nanomagnet** | Size | 20 nm X 30 nm | | |
| | Stability ($\Delta/kT$) | 40 | Magnetization ($M_s$) | < 500 MA/cm |
| | Spin injection | > 80 % | Spin polarization | > 80 % |

*Table 1. (Simulation parameters for transient vector spin simulation of MESO Logic device) Device and Material targets to enable 10 aJ class MESO logic are shown with references to demonstrated experimental values.*

|  | Type 1 | Type 2 | Type 3 |
|---|---|---|---|
| **SOC Materials For Spin to Charge** | Metals and Super-lattices | Topological Materials and Super-lattices | 2D materials |
|  | Bi/Ag[14], Pt[13], β-Ta[13], β-W[57], W(O)[71], CuBi, CuIr, $Bi_2O_3$[72], Ag/Sb[53] | $Bi_{1.5}Sb_{0.5}Te_{1.7}Se_{1.3}$[16], $Bi_2Se_3$[56], Bi-$Bi_2Se_3$[73], α-Sn[63] | $MoS_2$[50], $MX_2$[80] |
| **Magneto-electrics** | Multiferroics | Magneto-Strictive | Exchange Bias |
|  | $BiFeO_3$ [18], $LaBiFeO_3$[65], $BiCeFeO_3$[74], $LuFeO_3$[70] | $Fe_3Ga$[75], $Tb_xDy_{1-x}Fe_2$[76], FeRh[77] | $Cr_2O_3$[20,78], $Fe_2TeO_6$[79] |
| **Interconnect** | Noble Metals | Metal-Semiconductor | Interlayer Dielectric |
|  | Cu, Ag, Co, Al, Ru | poly-Si, NiSi, CoSi, NiGe, TiSi | $SiO_2$, SiN, SiCOH, Polymers |
| **Nanomagnet** | Nominal Ferro-magnets | Huesler Alloys |  |
|  | Co, Fe, Ni, CoFe, NiFe | $X_2YZ$, XYZ alloys e.g $Co_2FeAl$, $Mn_3Ga$ [59] |  |

*Table 2. Material Options for MESO logic.* Three classes of materials (Metals and superlattices, topological materials and superlattices, 2D transition Di Chalcogenides) are suitable for SOC based spin to charge conversion. Magneto-electrics also belong to three classes (Multiferroics with magnetic (AFM/FM) and electric (AFE/FE) order parameters, Magnetostrictive (one order FM order parameter material combined with a mechanical order parameter (piezoelectric), Exchange bias materials (one order parameter FM/AFM with no FE/AFE order). Magnetostrictive materials are not a direct suitable candidate (since only 90° switching is feasible) but can be used to augment the ME switching. Interconnect options comprise of Noble metals, Metal-semiconductors which exhibit excellent gap fill for interconnect processing and have short electron mean free path. Interlayer dielectrics are chosen for low refractive index for capacitance gains. Nanomagnets should be metallic to allow spin injection with applied bias. Co, Fe, Ni based or Heusler alloys are potential candidates with low $M_s$ and high spin polarization.

# Supplementary Materials

# Spin-Orbit Logic with Magnetoelectric Switching: A Multi-generation Scalable Charge Mediated Nonvolatile Spintronic Logic

Sasikanth Manipatruni, Dmitri E. Nikonov, Ramesh Ramamoorthy, Huichu Liu, Ian A. Young

**A. Vector Spin Circuit Modeling of MESO Logic:**

We verify the functionality of MESO spin logic using an equivalent spin circuit model, which comprehends magnetization dynamics of the nanomagnet, stochastic nature of switching, vector spin injection, spin to charge transduction and the magneto-electric switching. The equivalent circuit model is based on vector spin circuit theory [S1-4] (magneto-electric circuit analysis).

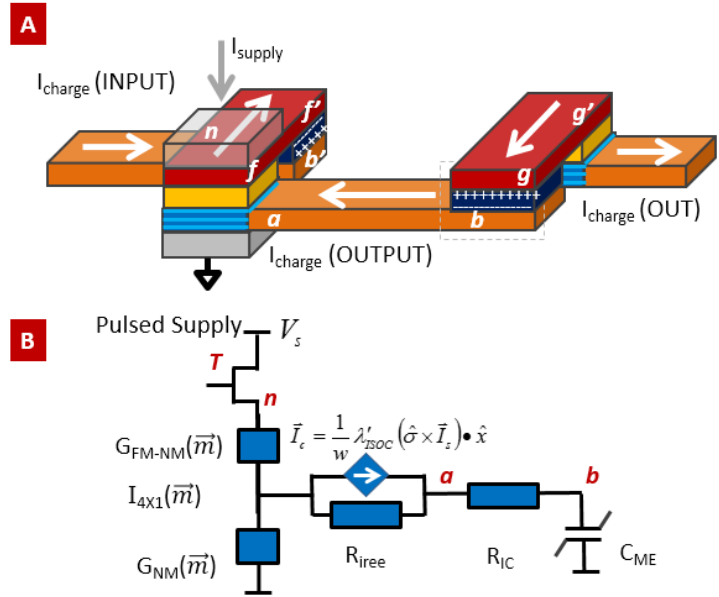

*Figure S1. Vector Spin Equivalent Model for MESO Logic device. A. Physical scheme. B. Circuit schematics.*

In figure S1, the node labels of the MESO circuit schematics correspond to the equivalent points in the physical scheme of the device. **ab** is a charge interconnect connecting two stages of MESO devices, where *a* is the node at the output terminal of the ISOC current source, *b* is the node at the terminal of the magneto-electric capacitor. **FF'** represents the ferromagnet (FM). The ferromagnet is adjacent to a magneto-electric capacitor **F'b'**. The spin to charge conversion layer is represented by a current controlled current source and an internal shunt resistance. Drive transistor (**T**) is connected to the FM to provide supply voltage. The pulse of current through the transistor is conducted to the ground terminal designated in the scheme by a triangle. Transistor (**T**) is shared among several MESO logic devices to provide supply voltage/current which plays the role of clocking. No logic signal is passed through these transistors. The ferroelectric capacitance of **bg** is designated $C_{ME}$. $G_{FM}$ is the 4x4 matrix of spin conductance of the interface between FM and the spin injection layer. $I_c$ is the equivalent current source due to spin to charge conversion in the SOC layer with intrinsic source resistance. Interconnect resistance is designated $R_{ic}$.

The functional form of the spin matrices $G_{FM}$ is obtained by using a Landauer-Büttiker formalism with spin transport [1, 5] applied to metallic circuits. The spin equivalent circuit modeling is described in [4]. The conductance matrix describing the spin transport across a ferromagnet (FM) to normal magnet (NM) interface is

$$\begin{bmatrix} I_c \\ I_{sx} \\ I_{sy} \\ I_{sz} \end{bmatrix} = \begin{bmatrix} G_{11} & \alpha(V_c)G_{11} & 0 & 0 \\ \alpha(V_c)G_{11} & G_{11} & 0 & 0 \\ 0 & 0 & G_{SL}(V_c) & G_{FL}(V_c) \\ 0 & 0 & -G_{FL}(V_c) & G_{SL}(V_c) \end{bmatrix} \begin{bmatrix} V_c \\ V_{sx} \\ V_{sy} \\ V_{sz} \end{bmatrix}$$

(S1)

where $G_{11}$ is the interface conductivity (for each interface), $\alpha(V)$ is the spin polarization of current across the interface as a function of voltage, $G_{SL}(V_c)$ and $G_{FL}(V_c)$ are the Slonczewski and field like torque contributions to the spin current across the interface. The voltage dependence of spin polarization $\alpha(V)$, $G_{SL}(V)$, $G_{FL}(V)$ is determined by the detailed band structure of the electrode materials (see e.g. [6, 7]). The expressions for the elements of the 4x4 spin conduction matrix (A1) have first been derived in [1]. The above expression is valid in a coordinate system where the x-axis is aligned to the direction of magnetization. For an arbitrary direction $\hat{m}$ of magnetization, the spin conduction matrix is obtained by applying a rotation transform

$$G_{FN}(\hat{m}) = R(\hat{m})^{-1} G_{FN}(\hat{x}) R(\hat{m}) \tag{S2}$$

Where the rotation matrix

$$R(\hat{m}) = \begin{bmatrix} 1 & 0 & 0 & 0 \\ 0 & r_{22} & r_{23} & r_{24} \\ 0 & r_{32} & r_{33} & r_{33} \\ 0 & r_{42} & r_{43} & r_{44} \end{bmatrix} \tag{S3}$$

The elements of $G_{FM}$ conductance can be related [1] to the ab-initio calculated reflection and transmission coefficients of the interface as follows

$$G^{\uparrow\uparrow} = \frac{e^2}{h} \sum_{m,n} \left| t_\uparrow^{nm} \right|^2 \tag{S4}$$

$$G^{\downarrow\downarrow} = \frac{e^2}{h} \sum_{m,n} \left| t_\downarrow^{nm} \right|^2 \tag{S5}$$

$$G^{\uparrow\downarrow} = G^{\downarrow\uparrow *} = \frac{e^2}{h} \sum_n \left( 1 - \sum_m r_\uparrow^{nm} r_\downarrow^{nm *} \right) \tag{S6}$$

where $e^2/h$ is the conductance per spin of a ballistic channel with ideal contacts [1]; $t_\downarrow^{nm}, t_\uparrow^{nm}$ are the transmission coefficients for majority and minority spin electrons; $r_\uparrow^{nm}, r_\downarrow^{nm}$ are the reflection coefficients of the majority and minority spin electrons; $n$ is the index of modes in the non-magnetic material, and *m* is the index of modes in the ferromagnet. From the above expressions we note relations

$$G_{SL} = G_{11} + \frac{e^2}{h}\sum_{m,n}\left|r_\uparrow^{nm} - r_\downarrow^{nm*}\right|^2 \qquad (S7)$$

$$G_{FL} = -i\frac{e^2}{h}\sum_{m,n}\left(r_\uparrow^{nm} r_\downarrow^{nm*} - r_\uparrow^{nm*} r_\downarrow^{nm}\right) \qquad (S8)$$

The functional form of the spin to charge conversion is simulated using a spin controlled current source (SCCS) modeled as a current source with open circuit resistance ($G_s$)

$$I_{c\,int} = \frac{1}{w}\left(\lambda_{IREE} + \Theta_{SHE}\lambda_{sf}\tanh\left(\frac{t}{2\lambda_{sf}}\right)\right)\left(\hat{\sigma}\times\vec{I}_s\right)\bullet\hat{y} \qquad (S9)$$

The bulk spin Hall coefficient may be enhanced by the use of a super-lattice of high spin orbit materials [9]. The interconnect resistance $R_{ic}$ and $C_{ic}$ are modeled by the following Mayadas-Shatzkes (MS) scaling law [10]

$$\rho = \rho_0\left(1 + \frac{3\lambda_{ebulk}}{8t}\left(1+\frac{p}{2}\right) + \frac{3\lambda_{ebulk}}{2D}\left(\frac{R}{1-R}\right)\right) \qquad (S10)$$

Where $\rho_0$ is bulk resistivity, $\lambda_{ebulk}$ is the electron mean free path, p, R are specularity, reflection parameter from grain boundaries) as critical interconnect dimensions approach electron mean free path. We assume that $\lambda_{ebulk} = p = R$. We also include the interconnect capacitance of the charge interconnect BC.

**B. Magnetization dynamics and stochastic behavior**

We modeled the magnetization dynamics of the nanomagnet using two complementary approaches: a) Landau-Lifshitz-Gilbert (LLG) equation [11] for the average magnetization dynamics affected by thermal noise; b) Fokker-Planck equation [12, 13] to determine the probability distribution of the stochastic part of magnetization. The phenomenological LLG equation describing the dynamics of nanomagnet with a magnetic moment unit vector ($\hat{m}$), modified with spin transfer torques is (see Table S1 for parameters)

$$\frac{\partial \hat{m}}{\partial t} = -\gamma\mu_0 [\hat{m} \times \vec{H}_{eff}] + -\gamma\mu_0 [\hat{m} \times \vec{H}_{exc}] + \alpha\left[\hat{m} \times \frac{\partial m}{\partial t}\right] + \frac{\vec{I}_\perp}{eN_s} \qquad (S11)$$

where γ is the electron gyromagnetic ratio; $\mu_0$ is the free space permeability; $\vec{H}_{eff}$ is the effective magnetic field due to the combination of material, shape, and surface anisotropy; α is the Gilbert damping and $\vec{I}_\perp$ is the component of spin polarization perpendicular to the magnetization ($\hat{m}$) in the current exiting the nanomagnet, $N_s$ is the total number of Bohr magnetons per magnet. $\vec{I}_\perp$ can be equivalently rewritten as $\vec{I}_\perp = \vec{I}_s - \hat{m}(\hat{m}.\vec{I}_s) = \hat{m} \times (\vec{I}_s \times \hat{m})$. The dynamics of nanomagnets is strongly affected by the thermal noise [14]. The thermal noise can be considered a result of microscopic degrees of freedom of the conduction electrons and the crystal lattice of the ferromagnet. At room temperature T, the thermal noise is described by a Gaussian white noise (with a time domain Dirac-delta auto-correlation). The noise field acts isotropically on the magnet. The internal field in (S12) is:

$$\vec{H}_{eff}(T) = \vec{H}_{eff} + (H_i\hat{x} + H_j\hat{y} + H_k\hat{z}) \qquad (S12)$$

$$\langle H_l(t) \rangle = 0 \qquad (S13)$$

$$\langle H_l(t)H_k(t') \rangle = \frac{2\alpha k_B T}{\mu_0^2 \gamma M_s V} \delta(t-t')\delta_{lk} \qquad (S14)$$

The initial conditions of the magnets should also be randomized to be consistent with the distribution of initial angles of magnet moments in a large collection of magnets.

At temperature T, the initial angle of the magnets follows:

$$\langle \theta^2 \rangle = \frac{kT}{M_s V \mu_0 H_{ani}} \quad (S15)$$

We used a mid-point integration method [15, 16] to apply the Stratonovich calculus while integrating the LLG equation. The discretized integration rule is

$$y_{n+1} = y_n + \Delta t . f(t_n + \Delta t/2, y_n + \Delta t/2 . f(t_n, y_n)) \quad (S16)$$

Where $y = d\hat{m}/dt$. The variance of the noise varies depending on the time step size. The discretization was performed internally using an implicit self-consistent solver.

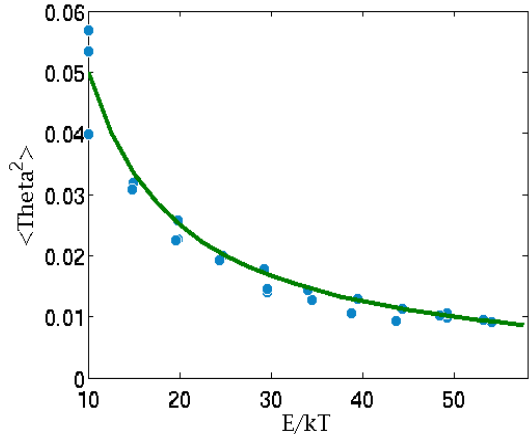

*Figure S2. Comparison of thermal noise induced variation in the free layer with thermodynamic model.*

### C. Logic Error Rates using Fokker-Planck equation

To study and compare the logic error rates (Figure 4D of main manuscript), we use Fokker-Planck equation governing the probability of the direction of the magnetic moment [12, 13, 17],

$$\frac{\partial \rho(\theta,\tau)}{\partial t} = -\nabla \cdot J(\theta,\tau) = -\frac{1}{\sin\theta}\frac{\partial}{\partial \theta}[\sin\theta J_\theta(\theta,\tau)] \quad \text{(S17)}$$

Which relates the rate of change of probability density $\rho(\theta,\tau)$ at a given angle to the net rate of probability density flow. $J_\theta(\theta,\tau)$ is the flow of probability given by drift and diffusion components

$$J_\theta(\theta,\tau) = \rho(\theta,\tau)\frac{\partial \theta}{\partial \tau} - D\frac{\partial \rho(\theta,\tau)}{\partial \tau} \quad \text{(S18)}$$

The flow term $\frac{\partial \theta}{\partial \tau} = (i - h - \cos\theta)\sin\theta$ where $i = I/I_c$, $h = H/H_{kc}$ are the current and field driving terms for the probability, D=1/2Δ is the diffusion constant given by the thermal barrier of the magnet. For completeness, we have compared the Fokker-Planck models with Monte Carlo simulations of the magnetization from (A12).

**D. Simulation Comparison of All Spin Logic and MESO Logic**

We compared the MESO logic with All Spin Logic (ASL) [18] using vector spin circuit modeling. All spin logic operates by injection of spin currents into the spin channels and the detection by spin torque. The assumed equivalent models for ASL devices are shown in figure S3. $G_{FM1}$ and $G_{FM2}$ are vector spin conductances describing the spin injection from the magnets to the spin channel. The spin conduction conductances $G_{SeT}$, $G_{sfT}$ describe the conduction through the spin channel. We represent the metallic spin channel between node 1 and 2 as a combination of two T-equivalent circuits. The conductance $G_{sfT}$ models the loss of spin current due to spin flip in the channel. The nanomagnet conductance is modeled by the spin conductance tensors $G_{FM1}(\hat{m}_1)$, $G_{FM2}(\hat{m}_2)$. The dynamics of the spin device are solved self-consistently with the spin transport in the equivalent circuit models. Please see [4, 19] for more details on the methodology.

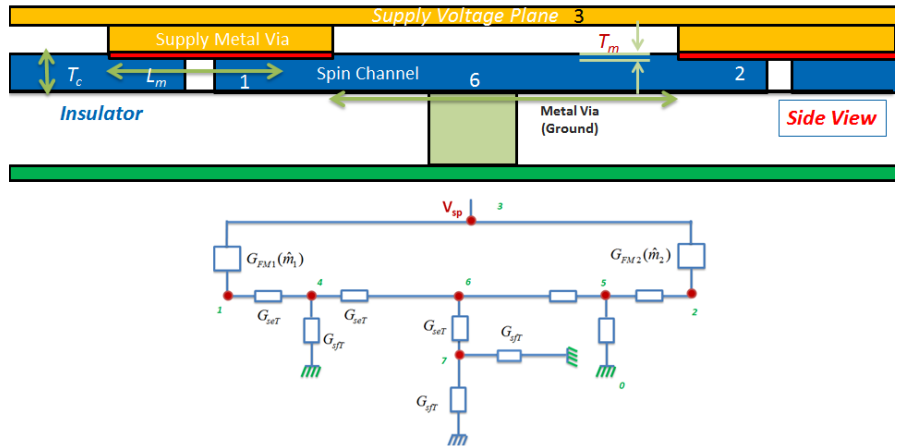

Figure S3. Benchmark All Spin Logic device and vector spin circuit model

| Table 1. Nanomagnet and transport parameters used for Spin torque logic | | | |
|---|---|---|---|
| **Variable** | **Notation** | **Value/Typical Value** | **Units (SI)** |
| Free Space Permeability | $\mu_0$ | $4\pi \times 10^{-7}$ | $JA^{-2}m^{-1}$ |
| Gyromagnetic ratio | $\gamma$ | $17.6 \times 10^{10}$ | $s^{-1}T^{-1}$ |
| Saturation Magnetization of the Magnet | $M_s$ | $10^6$ | A/m |
| Damping of the Magnet | $\alpha$ | 0.007 | - |
| Effective Internal Anisotropic Field | $H_{eff}$ | $3.06 \times 10^4$ | A/m |
| Barrier of the magnet | $\Delta/kT$ | 40 | |
| Number of spins | $N_s$ | $10^3 - 10^6$ | - |
| Thickness of Magnet | $T_m$ | 3 | nm |
| Magnet Dimensions | $W_m$ | 37.8X75.7 | nm |
| Length of channel | $L_c$ | 100 | nm |
| Thickness of channel | $T_c$ | 200 | nm |
| Length of ground lead | $L_g$ | 200 | nm |
| Thickness of ground lead | $T_g$ | 100 | nm |
| Channel conductivity | $\rho$ | $7 \times 10^{-9}$ | $\Omega.m$ |

| Sharvin conductivity | $G_{sh}$ | $0.5 \times 10^{15}$ | $\Omega^{-1} \cdot m^{-2}$ |
| --- | --- | --- | --- |
| Polarization | $\alpha_c$ | 0.8 | |

To benchmark the MESO device with ASL device, we considered two material systems: a) All Spin logic with perpendicular magnetic anisotropy (ASL-PMA) and perpendicular spin currents; b) All Spin logic with Heusler alloy PMA magnets (ASL-Heusler). The gate level energy delay of ASL-PMA and ASL-Heusler are 170 aJ·ns and 10 aJ·ns respectively [19].

For scaling the ASL devices, we assume the use of perpendicular anisotropy and Heusler alloys. The effect of scaling up the anisotropy field while reducing the saturation magnetization of the nanomagnets by the same factor allows strong improvement in ASL performance. The nanomagnet volume and the effective thermal energy barrier remain constant. A factor of 4 scaling of Ms and $H_k$ from the benchmark material (CoFeB) shows that the energy-delay product can reach 10 aJ·ns.

**E. Benchmarking to beyond CMOS logic options**

We adopted the benchmarking methodology [20, 21] for digital logic, applied to circuit such as a fanout-4 inverter, a 2-input NAND, and a 32-bit ripple-carry adders to compare the MESO logic with leading beyond CMOS logic options. Circuit layout of the terminal signals signal impacts the switching time/delay and energy. A detailed description of the layout assumptions (following the constraints of advanced lithography) are described in [20].

Numerical values in this study are obtained for the technology with the Dynamic Random Access Memory (DRAM) metal half-pitch F=15nm [22]. It is a measure of the semiconductor process's minimum critical dimension. Spintronic circuits are more compact due to the use of majority gates. Switching delay and energy of electronic devices are calculated according to well-known estimates for a capacitor charging. Switching estimates of a spintronic intrinsic element depends on the mechanism of switching: a) spin torque (for STT-DW: Spin Transfer Torque Domain Wall logic, ASLD: All Spin Logic Device, CSL: Charge Spin Logic, STO logic: Spin Torque Oscillator logic) b) dipole field (for NML, Nano-Magnetic Logic) c) magneto-electric (MESO, SMG: Spin Majority Gate logic, SWD: Spin Wave Device).

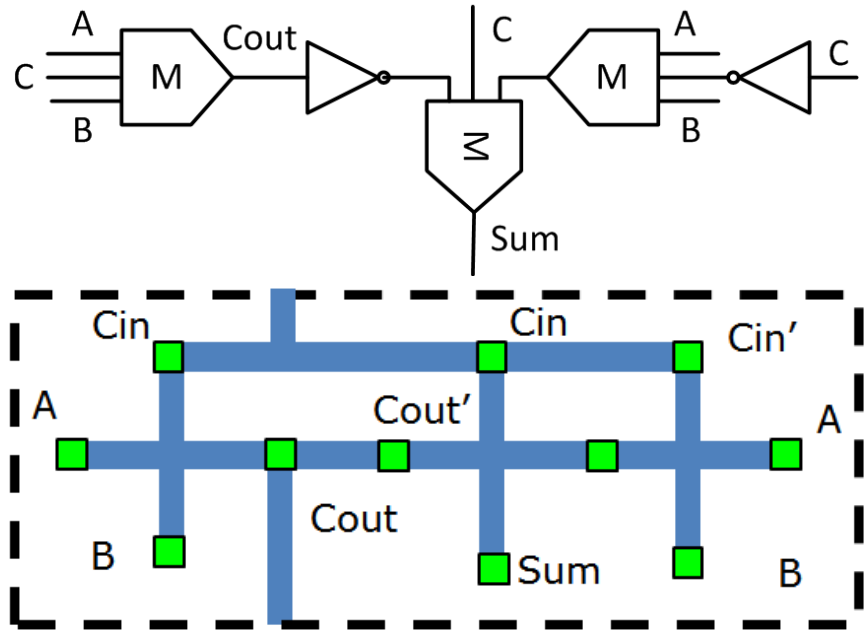

*Figure S4. Circuit schematics and layout of the benchmark circuit, a 32-bit adder (one bit cell of the adder is shown).*

To estimate the energy and the delay of the spin torque devices, we use the critical current $I_c = e\alpha\mu_0 M_s^2 v_{nm}/(\hbar P)$, the switching time $t_{stt} = eM_s v_{nm} L_f/(2g\mu_B P I_c)$, and the switching energy is $E_{stt} = 3I_c V_{dd} t_{stt}$. For devices switched by the magnetoelectric effect, the charging energy for a ME capacitor is $E_{me} = Q_{me} V_{dd}$, the magnetic switching time is limited by the ME field $t_{mag} = \pi / 2\gamma B_{me}$. Electronics is limited in supply voltage of 0.1 to 0.7V, while spintronics permits operation at 10 to 100mV, with the disadvantage of slower switching time. Both electronic and spintronic circuits need metal wires and contacts at the terminals of their gates. Switching time and energy of the intrinsic devices (transistors or nanomagnets), considering their parasitic as well as interconnects [23], are shown in figure 10.

### F. Energy calculation and scaling for MESO logic device

The switching energy ($E_{MESO}$) of a Magnetoelectric Spin-Orbit (MESO) logic device is composed of the sum of all the dissipation sources and energy storage: energy stored in the magnetoelectric ferroelectric (FE) ($E_{CME}$), dissipation in the interconnect ($E_{IC}$), dissipation in the spin to charge conversion layer ($E_{ISOC}$), dissipation in the power supply transistor ($E_{RT}$), dissipation in the supply and ground paths ($E_{SG}$).

$$E_{MESO} = E_{CME} + E_{IC} + E_{ISOC} + E_{RT} + E_{SG} \tag{S19}$$

### F.1. Equivalent charge circuit model for MESO logic device for energy calculation

MESO logic device comprises: a) magnetoelectric capacitor for voltage controlled switching of a ferromagnet (FM); b) a spin to charge conversion layer for charge readout of the magnetic state of the FM; and c) a charge interconnect connecting the MESO devices. An equivalent lumped element charge circuit model to capture the functioning of the MESO device is shown in Figure S5. Nodes a and b represent the two ends of the charge interconnect with interconnect resistance ($R_{ic}$). The spin to charge conversion element formed with inverse spin-orbit coupling (ISOC) materials is shown between node **n** and **a**. The ISOC module is modeled as a current controlled current source (CCCS) with an internal parallel resistance ($R_{isoc}$) [24, 25]. The charge dynamics of the magnetoelectric node are modeled via a ferroelectric capacitance ($C_{me}$) [26]. We present both an analytical expression for the total energy of a state transition of MESO, including the sum of all the dissipation and storage sources viz. energy stored in the ferroelectric (magnetoelectric), dissipation in the interconnect, dissipation in the supply-ground path, dissipation in the power supply transistor. We provide an analytical derivation of the transition energy, and support this with time-domain circuit simulations using a self-consistent ferroelectric compact model. The equivalent charge circuit is applicable for calculating the energy due to charge dynamics and does not capture the vector spin dynamics.

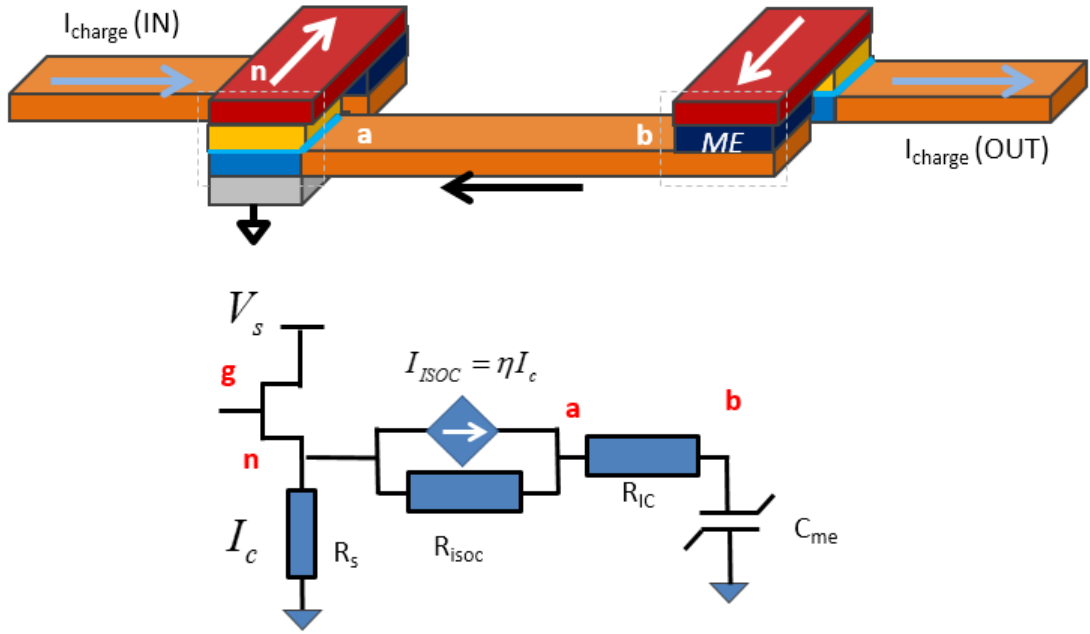

*Figure S5. **(Charge equivalent circuit for MESO)** MESO device comprising a magnetoelectric (ME) switchable capacitor representing the MESO input loading combined with a spin to charge transduction mechanism for the MESO magnetic state read-out. The spin to charge conversion is modeled as a current controlled current source with an internal resistance $R_{ISOC}$, the charge dynamics of the ME are modeled with a ferroelectric capacitor. $R_{ic}$ is the interconnect resistance forming the charge interconnect ab. $\eta$ is net spin to charge current conversion ratio.*

The explanation for the total energy of MESO device is as follows a) The energy to switch a capacitor and FE is independent of the interconnect resistance to the first order [24, 25] b) The current shunted in the spin to charge conversion current source depends on the equivalent source resistance ($R_{ISOC}$), which is material dependent parameter c) The losses in the parasitic paths are second order and device design can mitigate the energy losses extrinsic to magnetoelectric/ferroelectric switching. We also present an example method for mitigating energy losses extrinsic to magnetoelectric switching via proper choice of supply-ground path resistance ($R_s$).

### F.2. Energy calculation and scaling for MESO logic device
**a. Switching energy of a capacitor/ferroelectric is independent of the interconnect resistance**

We first note that the energy to switch a capacitor (or a fixed charge switchable device such as a ferroelectric) is independent of the interconnect resistance. For simplicity, we start with a linear dielectric capacitor. In the first stage it is switched from voltage 0 to voltage V through an interconnect of resistance of R by a voltage supply of V. We see that, the interconnect dissipated energy is independent of the interconnect resistance,

$$E_{Ohmic} = R\int_0^\infty i^2 dt = \frac{V^2}{R}\int_0^\infty e^{-2t/RC} dt = \frac{CV^2}{2} \quad (S20)$$

The total energy supplied by the voltage source is given by

$$E_{supply} = V\int_0^\infty idt = \frac{V^2}{R}\int_0^\infty e^{-t/RC}dt = CV^2 \qquad (S21)$$

So one half of it dissipated in the charging stage, according to S20, and the other half goes to the increase of energy of the capacitor. In the second stage, the voltage supply is removed, and the capacitor evolves from voltage V to voltage 0. In this stage the energy stored in the capacitor is dissipated in the resistor. Therefore, the overall energy dissipated in the charge-discharge cycle is equal to (S21).

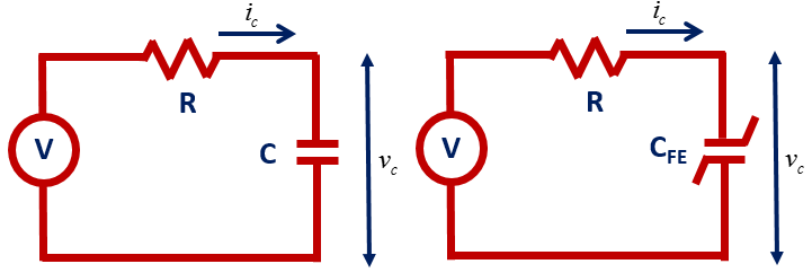

*Figure S6. RC circuit to show that the interconnect resistance does not impact the total energy to switch a capacitor. Linear dielectric capacitor used in the schematic on the left, a ferroelectric capacitor is used in the schematic on the right.*

The ferroelectric capacitor is treated somewhat differently. It starts with zero voltage but non-zero charge $-Q_{fe}$, corresponding to reversed spontaneous polarization $-P_{fe}$. It is then charged to voltage V and polarization $P_{fe}$. In the discharge stage, the voltage returns to zero, but there still remains the charge of $Q_{fe}$, corresponding to spontaneous polarization $P_{fe}$. The energy of the ferroelectric capacitor remains the same after the charge-discharge cycle. Similar to the linear dielectric capacitor, the total energy supplied by the voltage source to switch a ferroelectric is independent of the pulse shape of the current and is given by

$$E_{supply} = V\int_0^\infty idt = 2VQ_{fe} = C_{ME}V^2 \qquad (S22)$$

where we introduced the ferroelectric capacitance $C_{me}$. Hence, the Ohmic losses in switching a ferroelectric capacitor are only different from a linear capacitor by the factor of 2.

**b. Switching energy of MESO**

The total energy consumption of MESO can be written as sum of all the dissipation and storage sources Viz. energy stored in the FE, dissipation in the interconnect, dissipation in the supply-ground path, dissipation in the power supply transistor. We now consider two simplified equivalent models for the MESO for the energy consumption calculation comprising of the power supply (pulsed), Supply-ground path for spin to charge conversion, SOC current source, interconnect resistance and equivalent capacitance for ME. Figure S7A shows the simplified equivalent circuit to derive an analytical expression. We convert the spin to charge conversion current source (a current controlled current source) to a voltage source using Thevenin equivalence in Fig.S7 B.

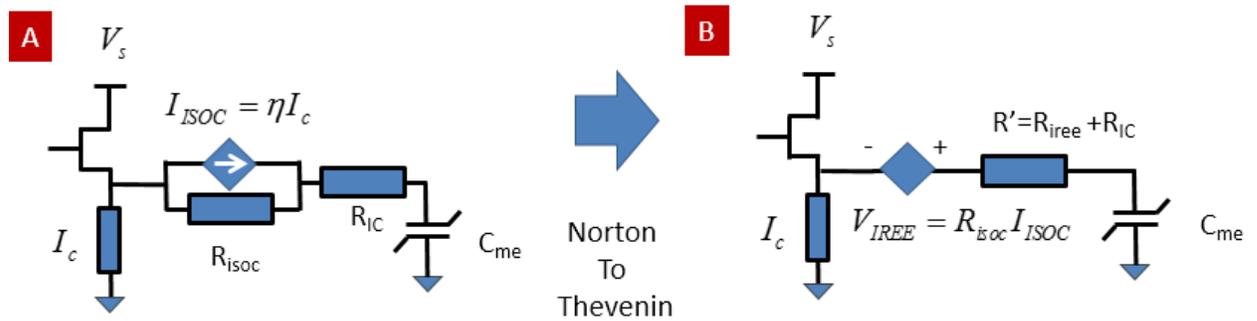

*Figure S7. A) Equivalent circuit for MESO with current controlled current source B) Current controlled voltage source. We apply Norton to Thevenin conversion.*

*Energy to switch the FE and the interconnect losses:* The total power dissipated in the interconnect resistance $R_{ic}$ and the ISOC internal resistance $R_{isoc}$ can be combined into R'. The power dissipated in the $R_{ic}$, $R_{isoc}$ and stored in ME capacitor are given by:

$$E_{CME} + E_{IC} + E_{ISOC} = C_{me}V_{me}^2 \qquad (S23)$$

*Joule losses occur in the resistances in the current delivery paths of the transistor and the resistive supply-ground path:* The energy dissipated in the supply-ground path is given by

$$E_{SG} = \int_0^\infty i_{sh}^2 R_{sh} dt = i_{sh}^2 R_{sh} \frac{2Q}{I_c} = kC_{me}V_{me}^2 \frac{W}{\lambda} \qquad (S24)$$

Where we use the relation, $i_{sh}R_{sh} = kV_{me}$, $i_{sh}/i_c = W/\lambda$, $2Q = C_{me}V_{me}$. W is the width of the magnet and λ is the ISOC parameter. It can be further shown that, the energy dissipated in the transistor and supply-ground path is given by

$$E_{SG} + E_{RT} = \alpha C_{me}V_{me}^2 \frac{W}{\lambda} \qquad (S25)$$

Where α is a circuit dependent function of transistor resistance, supply-ground path resistance.

The total energy of MESO can be written as

$$E_{MESO} = E_{CME} + E_{IC} + E_{ISOC} + E_{RT} + E_{SG} = C_{me}V_{me}^2\left(1 + \alpha\frac{W}{\lambda}\right) \qquad (S26)$$

### G. MESO parameters and benchmarks

Here we provide a detailed MESO switching energy calculation following the methodology of beyond-CMOS benchmarking [20, 21]. As the reader will see, the estimates below are in approximate agreement with the rigorous SPICE simulations (Section G). For our estimates we assume the following material and device parameters (Table 2). The calculation of the circuit area is outside the scope of this section. The charge circuit model of MESO operation is shown in Figure S5. Even though CMOS auxiliary circuits play a crucial role in the operation of MESO, here we will not describe the performance of a CMOS transistor. For details please see [20, 21].

I. <u>Magnetoelectric effect.</u>

In order to achieve the magnetoelectric field necessary to switch a nanomagnet, the following electric field needs to be applied to the multiferroic BiFeO$_3$:

$$E_r = E_{mf} B_c / B_{mf} . \tag{S27}$$

(For the definition of terms in Eq. (3.1) see Table 2 below). The total charge at the terminals of the multiferroic capacitor comprises the saturated ferroelectric polarization charge at the interface and the linear dielectric polarization in response to the applied electric field:

Table 2. Material and structure parameters serving as inputs into MESO estimates.

| Quantity | Symbol | Units | Value |
|---|---|---|---|
| Characteristic critical dimension | $F$ | m | 1e-8 |
| Copper wire resistivity | $\rho_{Cu}$ | $\Omega$*m | 2.5e-7 |
| Magnetization in a ferromagnet, perpendicular | $M_{sp}$ | A/m | 3e5 |
| Thickness of ferromagnetic | $t_{fm}$ | m | 2e-9 |
| Spin polarization from a ferromagnet | $P_{fm}$ | | 0.7 |
| Perpendicular magnetic anisotropy | $K_u$ | J/m$^3$ | 6e5 |
| Spin-orbit coupling effect coefficient | $\lambda_{ISOC}$ | m | 1.4e-8 |
| Resistance*area of the FM and ISOC stack | $r_{fm}$ | $\Omega$*m$^2$ | 3e-14 |

| Internal resistance of the ISOC current source | $R_{isoc}$ | Ω | 4000 |
|---|---|---|---|
| Magnetoelectric field for switching nanomagnet | $B_c$ | T | 0.1 |
| Multiferroic ferroelectric polarization (BFO) | $P_{mf}$ | C/m² | 0.3 |
| Multiferroic electric switching field | $E_{mf}$ | V/m | 1.8e6 |
| Multiferroic exchange bias at switching field | $B_{mf}$ | T | 0.03 |
| Dielectric constant of multiferroic | $\varepsilon_{mf}$ |  | 54 |
| Thickness of multiferroic | $t_{mf}$ | m | 5e-9 |
| Ferroelectric intrinsic switching time | $\tau_{fe}$ | s | 2e-11 |
| Lande factor | $g$ |  | 2 |
| Gate voltage for the access transistor | $V_x$ | V | 0.73 |
| On current for the access transistor | $\tilde{I}_x$ | A/m | 1648 |
| Gate capacitance per unit width of the access transistor | $c_g$ | F/m | 1e-9 |
| Resistance of a power or a ground distribution network | $R_s$ | Ω | 4000 |

$$Q_{me} = A_{me}\left(\varepsilon_0 \varepsilon_{mf} E_r + P_{mf}\right). \tag{S28}$$

The voltage drop on the magnetoelectric element is

$$V_{me} = E_r t_{mf}. \tag{S29}$$

The time to charge the multiferroic capacitor from 0 voltage to $V_{me}$ is

$$\tau_{me} \approx 2Q_{me} / I_{ISOC}, \tag{S30}$$

Since the charge in the capacitor needs to be changed from $-Q_{me}$ to $Q_{me}$. Here $I_{ISOC}$ is the current produced by the spin-orbit effect.

From the treatment below we will see that the capacitor charging time is limited also by the intrinsic switching time for the ferroelectric BFO to reverse polarization, $\tau_{fe}$. However the magnetization is even slower to react to the applied exchange bias and takes optimistically the following time (optimistically) to complete the precession:

$$\tau_{mag} \approx \pi/(\gamma B_c). \tag{S31}$$

Where the gyromagnetic ratio is

$$\gamma = ge/(2m_e). \tag{S32}$$

The total time is obtained as the combination of the above two times:

$$\tau_{tot} = \tau_{me} + \tau_{mag}, \tag{S33}$$

Where the last term makes the dominant contribution. The capacitance of the magnetoelectric element is

$$C_{me} = Q_{me}/V_{me}, \tag{S34}$$

And the switching energy is

$$E_{me} = 2Q_{me}V_{me}. \tag{S35}$$

### II. Spin-orbit coupling effect.

The power supply enable transistor provides current $I_c$ traveling through the ferromagnet. This current is related to the supply voltage $V_{sup}$ and the total resistance in the supply path.

$$V_s = I_c(R_T + R_{fm} + R_s) \equiv I_c R_{sum}. \tag{S36}$$

The enable transistor is in the linear regime. Its resistance is related to its width:

$$R_{ON} = r_L / w_x. \tag{S37}$$

This transistor width ($w_x$) is set by the objective to provide sufficient current to be converted by spin-orbit effect. The value of this width turns out to be smaller than a minimal transistor width. That means that one transistor can be shared to supply current to several parallel MESO devices, or a longer than minimum transistor channel length is used.

The resistance across the ferromagnet and spin-orbit coupling stack in the supply path is:

$$R_{fm} = r_{fm} / A_{me}. \tag{S38}$$

We assume that the contact resistance is included in the definition of the I-V characteristic of the MOSFET. $R_{ON}$ resistance is related to the on-current per unit width at the low source-to drain voltage and high gate to source voltage

$$r_L \approx V_x / \left(3\tilde{I}_x\right). \tag{S39}$$

The current extracts spin polarized current from the ferromagnet in the vertical direction

$$I_s = P_{fm} I_{sup}. \tag{S40}$$

Inverse spin-orbit coupling effect (the combination of the bulk spin Hall effect and the interface Rashba-Edelstein effect) converts the spin polarized current into charge current in the charging path (horizontal direction) of the multiferroic capacitor of the next MESO gate with its sign determined by the direction of magnetization in the ferromagnet.

$$I_{ISOC} = \lambda_{ISOC} I_s / w_m, \tag{S41}$$

Where $w_m$ is defined in table 1. This current source is related to the voltage it can produce, which must be equal to the magnetoelectric voltage:

$$V_{me} = I_{ISOC}(R_{ISOC} + R_{fe} + R_{ic}) = I_{ISOC}R_{tot}. \tag{S42}$$

The resistances in the charging path in the equation above are the ISOC current source internal resistance (longitudinal resistance of the thin ISOC layer), the resistance of the ferroelectric capacitor, and the interconnect resistance. The resistance of the ferroelectric capacitor is caused by damping in the ferroelectric and is related to the ferroelectric intrinsic switching time

$$R_{fe} = \tau_{fe}/C_{me}. \tag{S43}$$

The ISOC current source needs to be active over time $\tau_{me}$, necessary to charge the magnetoelectric capacitor. Thus energy dissipated in the vertical, supply path of the circuit is given by Joule power dissipation as:

$$E_{sup} = V_s I_c \tau_{me}. \tag{S44}$$

We can show that

$$E_{sup} = V_{sup} Q \frac{w_m}{P_{fm}\lambda_{IREE}}. \tag{S45}$$

Table 3. Operating parameters of MESO devices.

| Quantity | Symbol | Units | Value |
|---|---|---|---|
| Metal wire pitch (=4F) | $p_m$ | m | 4e-8 |
| Supply voltage | $V_s$ | V | 0.1 |
| Access transistor width | $w_x$ | m | 2e-9 |
| Resistance per width of the access transistor in the linear regime | $r_L$ | Ω*m | 1.48e-4 |

| Total resistance in the supply path | $R_{sum}$ | Ω | 2.5e4 |
|---|---|---|---|
| Characteristic interconnect length (=10p$_m$) | $l_{ic}$ | m | 4e-7 |
| Capacitance of a characteristic interconnect | $C_{ic}$ | F | 3.7e-17 |
| Resistance of a characteristic interconnect | $R_{ic}$ | Ω | 1e3 |
| Width of the magnet (=F) | $w_m$ | m | 1e-8 |
| Magnetoelectric switching area (=F$^2$) | $A_{me}$ | m$^2$ | 1e-16 |
| Magnetoelectric voltage | $V_{me}$ | V | 0.03 |
| Switching time for the magnetoelectric capacitor | $\tau_{me}$ | s | 5e-11 |
| Effective magnetoelectric capacitance | $C_{me}$ | F | 1e-15 |
| Resistance of the ferroelectric capacitor | $R_{fe}$ | Ω | 1e4 |
| Total resistance in the charging path | $R_{tot}$ | Ω | 1.8e4 |
| Switching time for magnetization | $\tau_{mag}$ | s | 2e-10 |
| Current in the supply path | $I_c$ | A | 1.2e-6 |
| Current generated by the ISOC effect | $I_{ISOC}$ | A | 1.2e-6 |
| Energy of magnetoelectric capacitor | $E_{me}$ | J | 1.8e-18 |
| Energy dissipation in the supply path | $E_{sup}$ | J | 6.4e-18 |
| Energy to charge the access transistor | $E_{ga}$ | J | 1.1e-18 |

Thus this contribution into energy dissipation is related to the energy of the magnetoelectric capacitor too. Together the energy loss in the C$_{me}$ charging/discharging and supply paths is

$$E_{MESO} = E_{me}\left(1 + \frac{w_m}{P_{fm}\lambda_{IREE}} \frac{V_s}{V_{me}}\right). \tag{S46}$$

An additional energy loss comes from charging the gate of the power supply gating transistor

$$E_{ga} = w_x c_g V_x^2. \tag{S47}$$

### III. MESO switching performance.

For a characteristic interconnect between MESO elements, the switching time is calculated according [20, 21, 27]:

$$t_{ic} \approx 0.38 R_{ic} C_{ic} + 0.7 R_{on} C_{ic} + 0.7 R_{ic} C_L, \tag{S48}$$

as well as the interconnect switching energy:

$$E_{ic} \approx C_{ic} V_{me}^2. \tag{S49}$$

Here one needs to substitute for the load capacitance $C_L = C_{me}$ and for the on-state resistance $R_{on} = (R_{ISOC} + R_{ic})$. The total delay time of an intrinsic device (not including the interconnect) is

$$t_{int} \approx \tau_{tot}, \tag{S50}$$

The intrinsic device in our case means 'one MESO element'. The way we calculate the performance of more complicated circuits, such as a 32-bit adder and an ALU, follows [20, 21]. Energy dissipation in the charging path of the circuit, is not a separate contribution to switching energy. It is equal to the energy accumulated in the magnetoelectric capacitor and just represents the way this energy is dissipated each time the voltage is turned on or off (see Section F for the explanation).

Thus the total intrinsic device switching energy is composed of

$$E_{int} = E_{me} + E_{sup} + E_{ga}. \tag{S51}$$

The results of the calculation following the method of [27] are summarized in Table 4.

Table 4. Resulting performance of MESO devices and circuits.

| Quantity | Symbol | Units | Value |
|---|---|---|---|
| Area of the intrinsic device | $a_{int}$ | m$^2$ | 1.4e-14 |
| Switching time of the intrinsic device | $t_{int}$ | s | 2.3e-10 |
| Switching energy of the intrinsic device | $E_{int}$ | J | 9.3e-18 |
| Switching time of the interconnect | $t_{ic}$ | s | 2.9e-12 |
| Switching energy of the interconnect | $E_{ic}$ | J | 1.8e-19 |
| Area of 1 bit of a full adder | $a_1$ | m$^2$ | 8.6e-14 |
| Switching time of 1 bit of a full adder | $t_1$ | s | 2.4e-10 |
| Switching energy of 1 bit of a full adder | $E_1$ | J | 1.3e-16 |

## H. SPICE simulation of the charge circuit with equivalent ferroelectric model

We validated our assumptions of the charge transport in the MESO device via a SPICE circuit simulation solver using compact model that comprehends the physics of the ferro-electrics and the spin to charge conversion. We model the ferroelectric switching dynamics of the magnetoelectric using Landau-Khalatnikov (LK) equation [26]:

$$\rho \frac{dQ_F}{dt} = -\frac{dU}{dQ_F} = -\left(2\alpha Q_F^1 + 4\beta Q_F^3 + 6\gamma Q_F^5 - V_F\right) \tag{S52}$$

Where $Q_F$ is the ferroelectric polarization, $\rho$ internal equivalent resistance (damping term) of the ferroelectric, U is the energy density per unit area, α, β, γ are the rescaled internal anisotropy constants of the FE. The ferroelectric switching exhibits a non-linear equivalent capacitance

$$C_F(Q_F) = \left(2\alpha + 4\beta Q_F^2 + 6\gamma Q_F^4\right)^{-1} \qquad (S53)$$

during the charging and discharging of the ferroelectric.

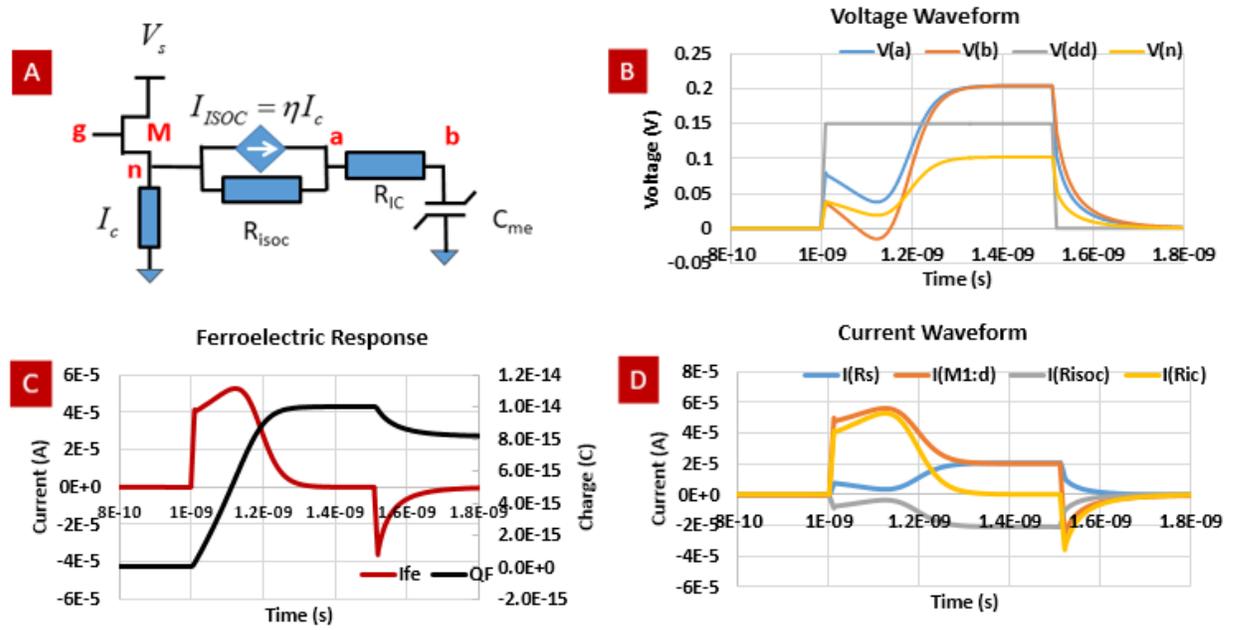

*Figure S8. SPICE Simulation of MESO showing the self-consistent charge dynamics of the MESO device **(for clarity of transient waveforms we simulated a 8 fC stored ferroelectric charge which is significantly higher than a scaled MESO logic device)** (A) Charge circuit model for MESO with ISOC current controlled current source and ME capacitor modeled with Landau-Khalatnikov equations B)Voltages applied/measured at supply transistor gate (V(dd)), magnetoelectric capacitor (V(b)), interconnect (V(a)) and the transistor terminal (V(n)) C) Current and charge across FE capacitor D) Currents measured through*

*ISOC internal resistance ($R_{isoc}$), supply resistance ($R_s$), supply transistor (M) & interconnect resistance ($R_{IC}$) ( Pulse width=500ps, $R_{isoc}=R_s=5k$, Vdd=150mV (clk), Vg=1V (dc))*

We perform time domain self-consistent SPICE simulations comprehending the ISOC current controlled current source with the non-linear dynamics of the Ferroelectric. The typical switching dynamics are shown in Figure S8. The supply is turned on from 1 ns to 1.5 ns. The switching dynamics at node **b** (FE capacitor terminal) are consistent with a ferroelectric switching via internal polarization dynamics. The charge and current through the FE capacitor is shown in Figure S8C. The current is consistent with classic FE switching pulse followed by a non-polarizing pulse at the end of the applied voltage.

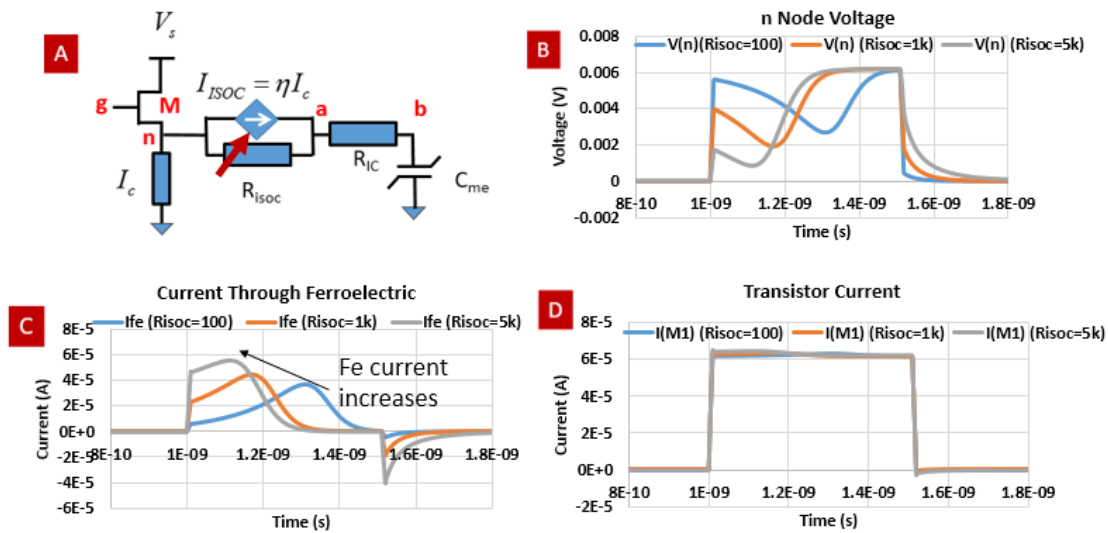

*Figure S9. **(Impact of ISOC internal current source resistance) (Impact of ISOC source resistance) (For clarity of transient waveforms we simulated an 8 fC stored ferroelectric charge which is significantly higher than a scaled MESO logic device)** SPICE Simulation of MESO showing the self-consistent charge dynamics of the MESO device B) Node voltage*

*at n1 at varying values of $R_{isoc}$ (100 Ω, 1kΩ, 5kΩ) C) Ferroelectric node current (Sweep $R_{isoc}$, Pulse width=500ps, $R_s$=100, Vdd=150mV, Vg=1V). D) Transistor current at varying values of $R_{isoc}$ (100 Ω, 1kΩ, 5kΩ) For high ISOC resistance FE node current approaches the transistor current for high spin to charge conversion ratio.*

We show the impact of the ISOC internal resistance and in particular show that interconnect current is close to the supply current for high internal resistance SOC materials. The currents measured through the various branches show that ISOC generated charge current is shared between the internal shunt path resistance and the interconnect resistance. The extent of shunting via internal resistance depends on the resistivity of the ISOC material (Figure S9).

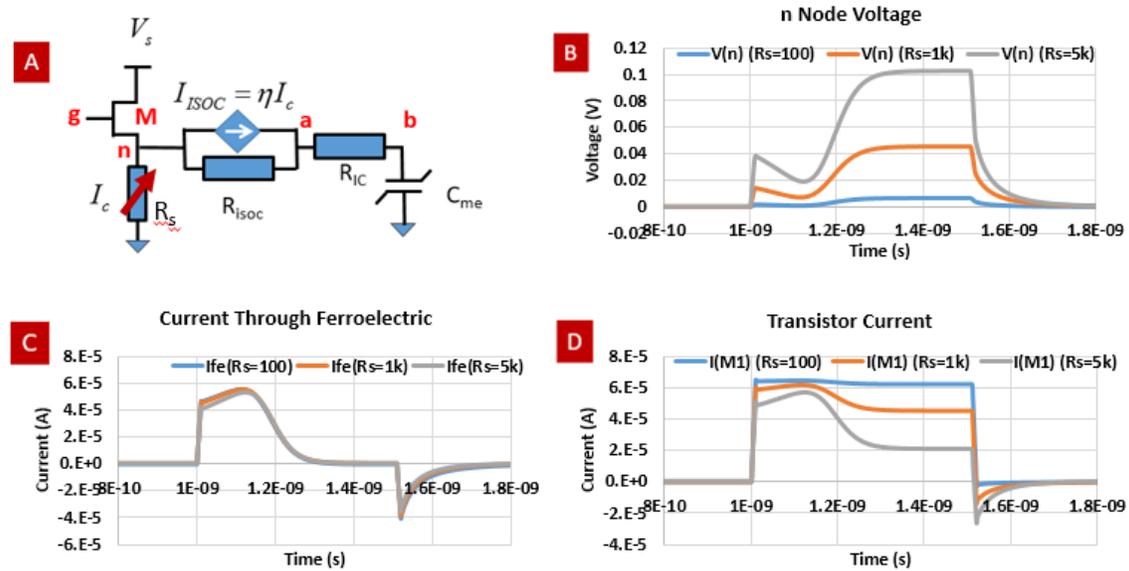

*Figure S10. **(Impact of $R_s$, Supply Resistance) (For clarity of transient waveforms we simulated an 8 fC stored ferroelectric charge which is significantly higher than a scaled MESO logic device)** A) SPICE Simulation of MESO showing the impact of $R_s$, supply resistance B) Node voltage at n at varying $R_s$ (100 Ω, 1kΩ, 5kΩ) C) Ferro-*

*electric/Interconnect current through $R_{IC}$ D) Transistor current at varying $R_s$ (100 Ω, 1kΩ, 5kΩ) (Sweep $R_s$, Pulse width=500ps, Risoc=5k, Vdd=150mV (clk), Vg=1V(dc))*

We study the impact of the supply resistor ($R_s$) on the switching dynamics of MESO. We show that the impact of supply path current (i.e., current through the resistor $R_s$) can be mitigated by a higher impedance without impacting the interconnect current and the switching dynamics of the ferroelectric (Figure S10).

**I. Scaled MESO switching operation comprehending all parasitic effects.**

We further show the SPICE simulation for a sub 10 aJ switching operation of a MESO logic device comprehending: a) Energy stored in $C_{me}$ b) Energy dissipation in the Interconnect-$R_{IC}$; c) Energy dissipation in the Rashba source's resistance $R_{isoc}$; d) Supply resistor losses.

The total energy of MESO is smaller than 10 aJ for scaled device dimensions with scaled material properties. Figure S11. shows the charge dynamics of a scaled MESO with ~35 aC stored charge corresponding to 35 µC.cm$^{-2}$ FE polarization charge density on a 10 nm X 10 nm ME capacitor. The impact of the supply current path can be mitigated via use of high resistive path to limit the current to the required magnitude for switching (Figure S11.B). A supply voltage of 100 mV is applied for 15 ps in Figure S11A, and the resultant voltages at magnetoelectric capacitor (**V(b)**), interconnect (**V(a)**) and the transistor terminal (**V(n)**) are shown. Currents measured through ISOC internal resistance ($R_{isoc}$), supply resistance ($R_s$), supply transistor (M) & interconnect resistance (RIC) are shown in Figure S11B. FE polarization charge with retention and charge storage of ~ 35 aC can be observed in Figure S11C. The current in the interconnect follows the voltage difference across the transistor node and the FE node.

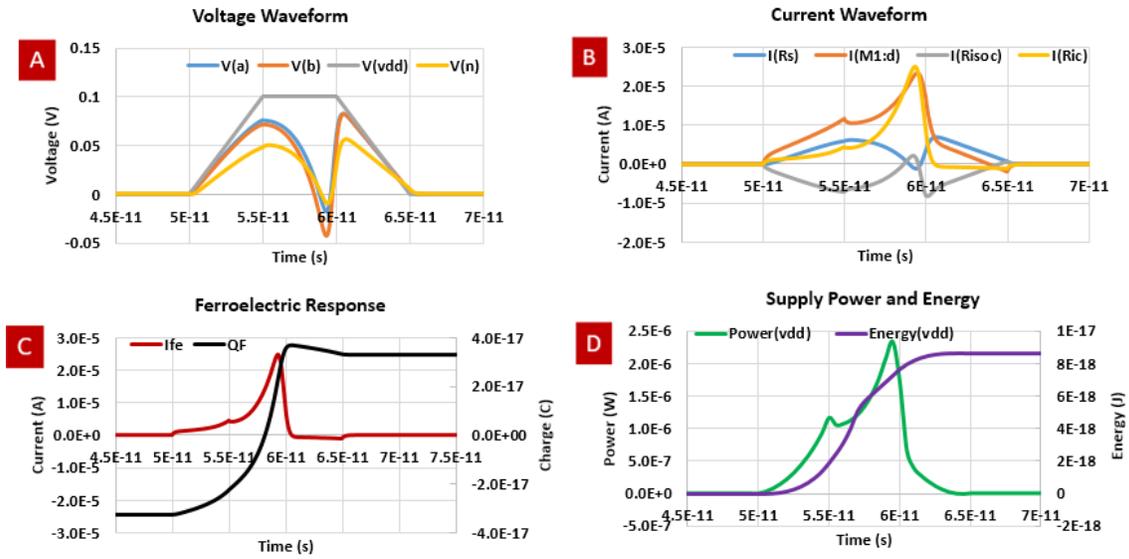

*Figure S11. **(Scaled MESO with 35 aC stored charge, sub 10 aJ switching energy)** A) Voltages applied/measured at supply transistor gate (**V(dd)**), magnetoelectric capacitor (**V(b)**), interconnect (**V(a)**) and the transistor terminal (**V(n)**) B) Currents measured through ISOC internal resistance ($R_{isoc}$), supply resistance ($R_s$), supply transistor (M) & interconnect resistance ($R_{IC}$) C) Ferroelectric node current and stored charge D) Supply power and energy vs time showing the total power delivered by the supply (Pulse width=15 ps, $R_{isoc}$=4 kΩ, $R_s$=8 kΩ, $R_{ic}$=1 kΩ, $V_{dd}$=100 mV (clk), $V_g$=0.8 V(dc))*

We probe the total power delivered by the supply for the scaled MESO to comprehend the effect of energy loss/storage from all the parts of the device. Figure.S11D shows that the total integrated energy per switching transistion is ~ 10 aJ. We note that these devices assumed ~ 35 aC of stored polarization per device which is equivalent to 35 µC.cm$^{-2}$. Further significant scaling in total energy is possible with a reduction of the FE polarization.

## J. Requirements for the resistivity of the spin-orbit coupling materials - Note on internal resistance of ISOC current controlled current source.

The output resistance of ISOC Current Controlled Current Source (CCCS) is obtained by dividing the open circuit voltage of the CCCS divided by the short circuit current [2, 3].The CCCS current source can also be converted to a current controlled voltage source (CCVS) by performing a Norton to Thevenin source conversion.

The ability of a current source to provide a current under resistive loading condition is improved as the internal resistance is increased. Research in spin-orbit coupling materials is opening up the possibility of high spin-orbit coupling materials with high intrinsic resistivity [28, 29], approaching 4-10 mΩ.cm. For example, a resistivity of 10 mΩ.cm [30] will provide an internal resistance of 5 kΩ - 20 kΩ for an ISOC spin to charge conversion layer with dimensions of 20 nm X 10-20 nm X 10-20 nm.

**References for Supplementary Material:**